\documentclass[
twocolumn,
showpacs,
preprintnumbers,
amsmath,
amssymb
]{revtex4-1}

\usepackage{graphicx,latexsym}
\usepackage{amsmath,amssymb,amsfonts}
\usepackage{bm}
\usepackage{braket}
\usepackage{mathrsfs}

\newcommand{\m}{\mathcal}

\usepackage{color}
\usepackage[dvipsnames]{xcolor}
\begin{document}
\title{Nonequilibrium Electron Distribution Function in a Voltage-Biased Nanowire:\\ A Nonequilibrium Green's Function Approach}
\author{Taira Kawamura$^1$ and Yusuke Kato$^{2,3}$}
\affiliation{$^1$Department of Physics, College of Science and Technology, Nihon University, Tokyo 101-8308, Japan}
\affiliation{$^2$Department of Basic Science, The University of Tokyo, 3-8-1 Komaba, Tokyo 153-8902, Japan}
\affiliation{$^3$Department of Physics, Graduate School of Science, The University of Tokyo, 7-3-1 Hongo, Tokyo 113-0033, Japan}
\date{\today}
%%%%%%%%%%%%%%%%%%%%%%%%%%%%%%%%%%%%%%%%%%%%%

\begin{abstract}
We develop a theoretical framework to determine distribution functions in nonequilibrium systems coupled to equilibrium reservoirs, by using the nonequilibrium Green's function technique. As a paradigmatic example, we consider the nonequilibrium distribution function in a nanowire under a bias voltage. We model the system as a tight-binding chain connected to reservoirs with different electrochemical potentials at both ends. For electron scattering processes in the wire, we consider both elastic scattering from impurities and inelastic scattering from phonons within the self-consistent Born approximation. We demonstrate that the nonequilibrium distribution functions, as well as the electrostatic potential profiles, in various scattering regimes are well described within our framework. This scheme will contribute to advancing our understanding of quantum many-body phenomena driven by nonequilibrium distribution functions that have different functional forms from the equilibrium ones.
\end{abstract}

\maketitle
%%%%%%%%%%%%%%%%%%%%%%%%%%%%%%%%%%%%%%%%%%%%%%%%%%%%%%%%%%%%%%%%%%%%
\section{Introduction}

Recent advances in experimental techniques for probing and controlling quantum many-body systems have stimulated theoretical interest in their nonequilibrium properties~\cite{Goldman2014, Bukov2015, Eckardt2015, Oka2019, Yin2022, Harper2020, Oka2009, Kitagawa2010, Lindner2011, Ezawa2013, Katan2013, Cayssol2013, Roy2017, Rudner2020, El2018, Yamamoto2019, Hanai2019, Hanai2020, Fruchart2021, Shen2018, Gong2018, Kawabata2019, Ashida2020, Borgnia2020, Bergholtz2021, Okuma2023, Wu2020}. In particular, periodically driven Floquet systems~\cite{Goldman2014, Bukov2015, Eckardt2015, Oka2019, Yin2022, Harper2020, Oka2009, Kitagawa2010, Lindner2011, Ezawa2013, Katan2013, Cayssol2013, Roy2017, Rudner2020} and open systems governed by non-Hermitian Hamiltonians~\cite{El2018, Yamamoto2019, Hanai2019, Hanai2020, Fruchart2021, Shen2018, Gong2018, Kawabata2019, Ashida2020, Borgnia2020, Bergholtz2021, Okuma2023} have attracted considerable attention due to their potential for realizing exotic quantum many-body states that have not been observed in thermal equilibrium systems. As exemplified by the Floquet/non-Hermitian topological band theories~\cite{Oka2009, Kitagawa2010, Lindner2011, Ezawa2013, Katan2013, Cayssol2013, Roy2017, Rudner2020, Shen2018, Gong2018, Kawabata2019, Ashida2020, Borgnia2020, Bergholtz2021, Okuma2023, Wu2020}, most theoretical studies of these systems have focused on nonequilibrium effects on their spectral properties, and the distribution functions describing their occupied states are assumed to follow the equilibrium forms, such as the Fermi-Dirac distribution function $f(\omega)=[1+e^{(\omega-\mu_{\rm eff})/T_{\rm eff}}]^{-1}$ characterized by effective ``temperature" $T_{\rm eff}$ and ``chemical potential" $\mu_{\rm eff}$.

These effective parameters are physically meaningful quantities when systems are in local equilibrium, where the distribution function $f^{\rm neq}_x(\omega)$ at each position $x$ is well fitted by the equilibrium distribution functions~\cite{Casas1994, Casas2003}. However, the fundamental differences in physical behavior between nonequilibrium quantum systems and their equilibrium counterparts emerge when the distribution function $f^{\rm neq}_x(\omega)$ deviates significantly from the equilibrium forms. While effective temperature and chemical potential are ill-defined in such highly nonequilibrium states, the distribution function $f^{\rm neq}_x(\omega)$ remains well-defined and serves as a useful quantity to characterize the nonequilibrium properties of systems. Thus, developing a theoretical framework to determine nonequilibrium distribution functions is crucial for exploring nonequilibrium quantum many-body phenomena beyond the local equilibrium paradigm.

\begin{figure}[t]
\centering
\includegraphics[width=8.2cm]{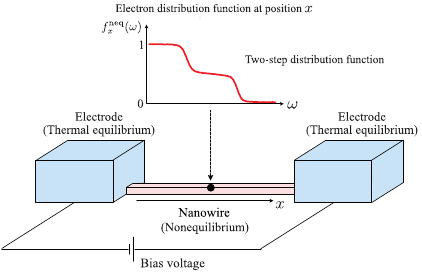}
\caption{A nanowire connected between two electrodes with different electrochemical potentials due to the bias voltage. These electrodes can be approximated as isolated systems in thermal equilibrium, where electrons follow the Fermi-Dirac distribution function. On the other hand, electrons at position $x$ in the wire follow a nonequilibrium distribution function $f^{\rm neq}_x(\omega)$, which in general has a different functional form from the Fermi-Dirac distribution function. The form of the nonequilibrium distribution function $f^{\rm neq}_x(\omega)$ depends on scattering processes experienced by electrons as they traverse the wire~\cite{Pothier1997, GueronThesis, Anthore2003, Huard2005, HuardThesis,  PierreThesis, AnneThesis, Tikhonov2020, Franceschi2002, Chen2009, Bronn2013}. When the wire length is shorter than the electron inelastic mean free path, $f^{\rm neq}_x(\omega)$ exhibits the two-step structure at low temperatures, reflecting the different electrochemical potentials in the electrodes.}
\label{fig.wire}
\end{figure}

A nonequilibrium system coupled to equilibrium reservoirs reaches a nonequilibrium steady state (NESS) through the balance between driving forces and dissipations. The nonequilibrium distribution function in the NESS is determined by solving a boundary value problem, with the equilibrium distribution functions in the reservoirs serving as boundary conditions. A notable example of such nonequilibrium distribution functions is the ``two-step distribution function" observed in mesoscopic systems under bias voltage~\cite{Pothier1997, GueronThesis, Anthore2003, Huard2005, HuardThesis,  PierreThesis, AnneThesis, Tikhonov2020, Franceschi2002, Chen2009, Bronn2013, Bronn2013_2, Stegmann2018}. In a metal wire between two reservoirs (electrodes) with different electrochemical potentials, electrons follow a position-dependent nonequilibrium distribution function $f^{\rm neq}_x(\omega)$, as schematically illustrated in Fig.~\ref{fig.wire}. This distribution function has been experimentally observed by superconducting tunneling spectroscopy~\cite{Pothier1997, GueronThesis, Anthore2003, Huard2005, HuardThesis,  PierreThesis, AnneThesis}, shot noise measurements~\cite{Tikhonov2020}, and using the Kondo effect in quantum dot systems~\cite{Franceschi2002}. In particular, when the wire length is shorter than the electron inelastic mean free path, a distribution function with a two-step structure emerges at low temperatures, reflecting the Fermi-Dirac distribution functions in the electrodes that have different electrochemical potentials (see Fig.~\ref{fig.wire}). Similar two-step distribution functions have been observed in voltage-biased carbon nanotubes~\cite{Chen2009, Bronn2013, Bronn2013_2}, and their potential realization in ultracold Fermi gases in a two-terminal configuration has also been explored~\cite{Lebrat2018, Mohan2024}.

%%%%%%%%%%%%%%%%%%%%%%%%%%%%%%%%
%%%%%%%%%%%%%%%%%%%%%%%%%%%%%%%%

Nonequilibrium distribution functions such as the two-step distribution function can lead to a variety of interesting phenomena~\cite{Baselmans1999, Shaikhaidarov2000, Baselmans2001, Baselmans2002, Huang2002, Pandey2022, Abanin2005, Kawamura2020, Kawamura2022, Kawamura2024, kawamura2024review, Kawamura2025, Dmitriev2003, Dmitriev2005, Dorozhkin2005, Dorozhkin2016, Clarke1972, Tinkham1972, Tinkham1972PRB, Schmid1975}. For instance, it has been experimentally demonstrated that the two-step distribution function can be used to control the critical current of a Josephson junction and realize $\pi$ junction~\cite{Baselmans1999, Shaikhaidarov2000, Baselmans2001, Baselmans2002, Huang2002, Pandey2022}. Moreover, the two-step distribution function can induce anomalous Fermi edge singularities~\cite{Abanin2005} and spatially inhomogeneous Fulde-Ferrell-Larkin-Ovchinnikov-type superconducting states~\cite{Kawamura2020, Kawamura2022, Kawamura2024, kawamura2024review, Kawamura2025}. Besides these phenomena associated with the two-step distribution function, in two-dimensional electron gases in semiconductor heterostructures exposed to microwave radiation, oscillatory structure in the distribution function is known to induce magnetoresistance oscillations~\cite{Dmitriev2003, Dmitriev2005, Dorozhkin2005, Dorozhkin2016}. In superconductors under quasiparticle injection, the nonequilibrium quasiparticle distribution generates the pair-quasiparticle potential difference, known as charge imbalance~\cite{Clarke1972, Tinkham1972, Tinkham1972PRB, Schmid1975}.

Moreover, nonequilibrium distribution functions also serve as a powerful tool for probing thermalization in quantum systems. In quantum Hall edge channels, a two-step distribution generated by a voltage-biased quantum dot is injected into an edge mode and measured after propagation~\cite{Altimiras2010, Altimiras2010_2, Sueur2010, Itoh2018}. Because the edge channel behaves as an ideal ballistic quantum channel, the injected two-step structure has been observed to propagate without thermalization. These examples demonstrate that a proper description of nonequilibrium distribution functions is crucial for understanding a wide range of nonequilibrium phenomena beyond the local-equilibrium paradigm.

In this paper, we develop a theoretical framework to determine position-dependent distribution functions $f^{\rm neq}_x(\omega)$ in nonequilibrium systems coupled to equilibrium reservoirs, by employing the nonequilibrium Green’s function technique~\cite{RammerBook, StefanucciBook, HaugBook, CamsariBook}. While the nonequilibrium Green's function technique has been widely used to study nonequilibrium quantum systems~\cite{RammerBook, StefanucciBook, HaugBook, CamsariBook}, its application to the boundary value problems for nonequilibrium distribution functions is very limited. As a paradigmatic example, we consider a nonequilibrium distribution function $f^{\rm neq}_x(\omega)$ in a voltage-biased nanowire illustrated in Fig.~\ref{fig.wire}. In this system, the electrodes connected to both ends of the wire can be approximated as reservoirs in thermal equilibrium, which serve as the boundary conditions for the nonequilibrium distribution function in the wire. The form of the distribution function $f^{\rm neq}_x(\omega)$ depends on scattering processes experienced by electrons as they traverse the wire~\cite{Pothier1997, GueronThesis, Anthore2003, Huard2005, HuardThesis,  PierreThesis, AnneThesis, Tikhonov2020, Franceschi2002, Chen2009, Bronn2013}. We consider elastic scattering from impurities, as well as inelastic scattering from phonons, and systematically investigate how these scattering processes affect the form of the distribution function $f^{\rm neq}_x(\omega)$.

We make a remark on the difference between the nonequilibrium Green's function approach and the transport equation approach used in previous work~\cite{Nagaev1992, Nagaev1995, Kozub1995, Naveh1998}. The  nonequilibrium (Wigner) distribution function $f^{\rm neq}_x(\bm{p})$ follows the Boltzmann equation~\cite{heikkilaBook}
\begin{equation}
\big[v_x \partial_x + e\bm{E}\cdot \partial_{\bm{p}}\big] f^{\rm neq}_x(\bm{p}) =
I_{\rm coll}\big\{f^{\rm neq}_x(\bm{p})\big\},
\label{eq.Boltzmann0}
\end{equation}
which describes the semiclassical motion of an electron with momentum $\bm{p}=m\bm{v}$  in the electric field $\bm{E}$. Here, we use the one-dimensional form, assuming homogeneity in the lateral directions. In Eq.~\eqref{eq.Boltzmann0}, $I_{\rm coll}$ is the collision term, which describes the electron scattering effects. In the case of strong impurity scattering (diffusive limit), the distribution function is almost isotropic in momentum $\bm{p}$ space, and it can be regarded as a function of the electron kinetic energy $\omega=\bm{p}^2/(2m)$. Averaging over momentum directions in Eq.~\eqref{eq.Boltzmann0}, one obtains the equation for the distribution function as~\cite{Nagaev1992, Nagaev1995, Kozub1995, Naveh1998, heikkilaBook}
\begin{equation}
D\partial_x^2 f^{\rm neq}_x(\omega) = I_{\rm inel}\{f^{\rm neq}_x(\omega)\},
\label{eq.Boltzmann.diffusive}
\end{equation}
where $D$ is the diffusion constant and $I_{\rm inel}$ describes the effects of inelastic electron scattering. In the previous works~\cite{Nagaev1992, Nagaev1995, Kozub1995, Naveh1998}, the nonequilibrium distribution function in the metal wire depicted in Fig.~\ref{fig.wire} is determined by solving Eq.~\eqref{eq.Boltzmann.diffusive} with boundary conditions
\begin{subequations} \label{all}
\begin{align}
& f^{\rm neq}_{x=0}(\omega)=f(\omega-\mu_{\rm L}),
\label{eq.BC1}
\\
& f^{\rm neq}_{x=L}(\omega)=f(\omega-\mu_{\rm R}),
\label{eq.BC2}
\end{align}
\end{subequations}
which are imposed by the reservoirs at both ends of the wire. Here, $L$ denotes the wire length and $f(\omega -\mu_{\alpha={\rm L}, {\rm R}})$ is the Fermi-Dirac distribution function in the left and right reservoir with the electrochemical potential $\mu_\alpha$. This approach, however, has a limitation. Since Eq.~\eqref{eq.Boltzmann.diffusive} is applicable only in the diffusive limit, we need to solve the more general Boltzmann equation~\eqref{eq.Boltzmann0} to deal with systems in the ballistic-diffusive crossover regime. However, we cannot impose the two boundary conditions, such as Eqs.~\eqref{eq.BC1} and \eqref{eq.BC2}, on the Boltzmann equation~\eqref{eq.Boltzmann0} because it is a first-order differential equation with respect to $x$. Thus, the applicability of the transport equation approach to boundary value problems for nonequilibrium distribution functions is restricted to systems in the diffusive limit. In contrast, the nonequilibrium Green’s function approach, which incorporates system-reservoir coupling effects through self-energy corrections, does not suffer from the difficulty of imposing boundary conditions. As a result, this approach enables a unified description of nonequilibrium distribution functions across the ballistic-diffusive crossover regime.

This paper is organized as follows. In Sec.~\ref{sec.formalism}, we present our model of a voltage-biased nanowire and explain how to determine the nonequilibrium distribution function in the wire by using the nonequilibrium Green's function technique. In Sec.~\ref{sec.result}, we show the calculated nonequilibrium distribution function and discuss electron scattering effects. Throughout this paper, we set $\hbar=k_{\rm B}=1$ and take $e<0$.

%%%%%%%%%%%%%%%%%%%%%%%%%%%%%%%%%%%%%%%%%%%%%%%%%%%%
%%%%%%%%%%%%%%%%%%%%%%%%%%%%%%%%%%%%%%%%%%%%%%%%%%%%
\section{Formalism \label{sec.formalism}}

%%%%%%%%%%%%%%%%%%%%%%%%%%%%%%%%
\subsection{Model \label{sec.model}}

\begin{figure}[t]
\centering
\includegraphics[width=8.2cm]{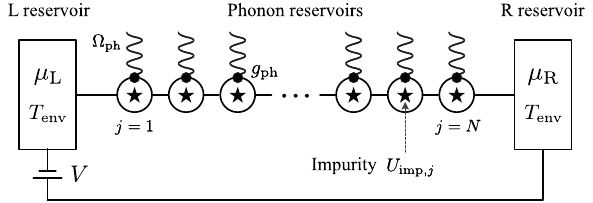}
\caption{Schematic picture of our model. Two free-fermion $\alpha$ ($={\rm L}, {\rm R}$) reservoirs are connected to both ends of the tight-binding chain with $N$ sites. The $\alpha$ reservoir is in the thermal equilibrium state characterized by the electrochemical potential $\mu_\alpha$ and the temperature $T_{\rm env}$. The potential difference $\mu_{\rm L}-\mu_{\rm R}$ equals the applied bias voltage $eV$ across the wire. Within the chain, electrons scatter from randomly distributed impurities with the potential $U_{{\rm imp}, j}$. The electrons also interact with local phonons of frequency $\Omega_{\rm ph}$, where $g_{\rm ph}$ represents the electron-phonon coupling constant.}
\label{fig.model}
\end{figure}

We consider a nanowire connected between two electrodes under a finite bias, as illustrated in Fig.~\ref{fig.wire}. To study the nonequilibrium electron distribution along the wire, we consider a tight-binding chain connected to two reservoirs with different electrochemical potentials, as illustrated in Fig.~\ref{fig.model}. The Hamiltonian is
\begin{equation}
H = H_0 + H_{\rm lead} + H_{\rm t} + H_{\rm imp} + H_{\rm ph} + H_{\rm e-ph}.
\label{eq.model.H}
\end{equation}
Here,
\begin{equation}
H_0 = -t \sum_{j=1}^{N-1} \big[c^\dagger_{j} c_{j+1} + {\rm H.c.}\big] + e \sum_{j=1}^N \varphi_j c^\dagger_{j} c_j
\label{eq.H0}
\end{equation}
is the one-dimensional tight-binding Hamiltonian used to model the nanowire. In Eq.~\eqref{eq.H0}, $N$ denotes the number of lattice sites, $-t$ is the nearest-neighbor hopping amplitude,  and $\varphi_j$ represents the electrostatic potential at site $j$ ($=1,\cdots,N$). For simplicity, we neglect spin-dependent interactions in this work, which allows us to treat the electrons as spinless. For later convenience, we define a parameter
\begin{equation}
x_j = \frac{j-1}{N-1},
\label{eq.xj}
\end{equation}
which specifies the distance from the left end of the wire.

The use of the single-channel model in Eq.~\eqref{eq.H0} is well justified when the transverse confinement of the wire is sufficiently strong, such that the energy spacing of the transverse modes exceeds both the thermal energy and the applied bias. Such conditions are typically realized in semiconductor quantum wires. 

In contrast, using the model in Eq.~\eqref{eq.H0} to describe a metallic wire requires more caution. Real metallic wires support a large number of conducting channels, and reducing them to a single effective mode constitutes a strong simplification. As a result, the strictly one-dimensional model employed here cannot capture phenomena that intrinsically rely on multichannel physics or on anisotropic deformation of the Fermi surface.

%%%%%%%%%%%%%%%%%%%%%%%%%%%%%%%%
%%%%%%%%%%%%%%%%%%%%%%%%%%%%%%%%

The electrodes connected to both ends of the tight-binding chain are described by $H_{\rm lead}$, having the form
\begin{equation}
H_{\rm lead} = \sum_{\alpha={\rm L}, {\rm R}} 
\sum_{\bm{k}} \xi_{\alpha,\bm{k}} a^\dagger_{\alpha,\bm{k}} a_{\alpha,\bm{k}}.
\end{equation}
Here, $a^\dagger_{\alpha, \bm{k}}$ creates an electron with kinetic energy $\xi_{\alpha,\bm{k}}$ in the $\alpha$ ($=$L, R) reservoir. The reservoirs are assumed to be in the thermal equilibrium state characterized by their electrochemical potential $\mu_{\alpha}$ and temperature $T_{\rm env}$. Under this assumption, electrons in the $\alpha$ reservoir follow the Fermi-Dirac distribution function,
\begin{equation}
f(\omega -\mu_\alpha) = \frac{1}{e^{(\omega-\mu_\alpha)/T_{\rm env}}+1}.
\end{equation}
The applied bias voltage $eV$ across the wire equals the electrochemical potential difference $\mu_{\rm L}-\mu_{\rm R}$ between the left and right reservoirs.

The coupling between the wire and the electrodes is described by
\begin{equation}
H_{\rm t}= -\sum_{\bm{k}}\big[t_{\rm L}a^\dagger_{{\rm L},\bm{k}} c_1 +{\rm H.c.} \big] -\sum_{\bm{k}}\big[ t_{\rm R} a^\dagger_{{\rm R},\bm{k}} c_N  +{\rm H.c.} \big].
\end{equation}
Here, $-t_\alpha$ is the hopping amplitude between the wire and the $\alpha$ reservoir. For simplicity, we consider the case of symmetric coupling ($t_{\rm L}=t_{\rm R}\equiv t_{\rm lead}$), which allows us to set $\mu_{\rm L}=+eV/2$ ($>0$) and $\mu_{\rm R}= -eV/2$.

The form of the electron distribution function reflects the scattering processes experienced by electrons as they traverse the wire. In this work, we examine how the distribution function is affected by two scattering processes: {\it elastic} scattering from (non-magnetic) impurities and {\it inelastic} scattering from phonons. The elastic scattering is described by $H_{\rm imp}$ in Eq.~\eqref{eq.model.H}, having the form
\begin{equation}
H_{\rm imp} = \sum_{j=1}^N U_{{\rm imp},j} c^\dagger_j c_j.
\end{equation}
Here, $U_{{\rm imp},j}$ represents the impurity scattering potential at site $j$, given by
\begin{equation}
U_{{\rm imp},j} = u_{\rm imp}\, n_{{\rm imp},j},
\end{equation}
where $n_{{\rm imp},j}$ $(=0,1)$ is a binary variable indicating the absence or presence of an impurity at site $j$. We decompose the impurity potential into its spatial average and the fluctuation around it,
\begin{equation}
U_{{\rm imp},j}
= \bar{U}_{\rm imp} + \delta U_j
= u_{\rm imp}\bigl[\bar{n}_{\rm imp} + \delta n_{{\rm imp},j}\bigr],
\label{eq.Uimp}
\end{equation}
where $\bar{n}_{\rm imp}= N_{\rm imp}/N$ denotes the average impurity concentration and $\delta n_{{\rm imp},j}= n_{{\rm imp},j} -\bar{n}_{\rm imp}$ the deviation from it. Since the average component $\bar{U}_{\rm imp}$ produces only a constant energy shift, we set $\bar{U}_{\rm imp}=0$.

The phonons are introduced as local harmonic oscillators at each site, known as the Holstein model in the literature~\cite{Holstein1959, Holstein1959_2, StefanucciBook}. The phonon reservoirs are described by
\begin{equation}
H_{\rm ph}= \sum_{j=1}^N \Omega_0 b^\dagger_j b_j,
\end{equation}
where $\Omega_0$ represents the phonon frequency and $b_j$ denotes the phonon annihilation operator at site $j$. The parameter $\Omega_0$ sets the characteristic energy scale for inelastic scattering processes, determining the typical energy quanta exchanged between electrons and phonons. The electron-phonon interaction in the wire is described by
\begin{equation}
H_{\rm e-ph}= g_{\rm ph} \sum_{j=1}^N a^\dagger_j a_j \big[b_j +b^\dagger_j\big],
\end{equation}
where $g_{\rm ph}$ represents the electron-phonon coupling constant. In this model, the strength of the electron-phonon coupling can be characterized by the parameter
\begin{equation}
\gamma_{\rm ph} \equiv 2g^2_{\rm ph}/\Omega_0,
\end{equation}
which gives the strength of the phonon-mediated on-site attractive interaction in the antiadiabatic limit. We note that this local Holstein-type model simplifies the spatial structure of the interaction and leads to a diagonal self-energy in site index. As we will demonstrate later, this simple model already captures the qualitative features of local thermalization induced by inelastic scattering~\cite{Note.nonlocal}.

It is worth noting that electron-electron scattering also affects the form of the distribution function~\cite{Pothier1997, Anthore2003, Huard2005, HuardThesis, GueronThesis, PierreThesis, AnneThesis, heikkilaBook}. It is known that in mesoscale ($\sim 1\mu$m) diffusive metal wire, electron-electron scattering due to screened Coulomb interactions dominates over electron-phonon scattering at low temperatures, typically below 1K~\cite{Altshuler1982, EfrosBook, heikkilaBook, Pierre2003}. Moreover, a tiny concentration of magnetic impurities with a small Kondo temperature enhances electron-electron scattering effects~\cite{Kaminski2001, Goppert2002, Pierre2003, Anthore2003, Huard2005, HuardThesis, PierreThesis, AnneThesis}. However, the theoretical treatment of these strong correlation effects is beyond the scope of this study.

%%%%%%%%%%%%%%%%%%%%%%%%%%%%%%%%
\subsection{Nonequilibrium Green's Function}

To determine the nonequilibrium distribution function in the nanowire, we conveniently introduce a $N\times N$ matrix nonequilibrium Green's function, given by
\begin{equation}
\bm{G}^{{\rm X}={\m{R}}, {\m{A}},\lessgtr}(t, t')
=
\begin{pmatrix}
G^{\rm X}_{11}(t,t') & \cdots & G^{\rm X}_{1N}(t,t') \\
\vdots & \ddots & \vdots \\
G^{\rm X}_{N1}(t,t') & \cdots & G^{\rm X}_{NN}(t,t') 
\end{pmatrix},
\label{eq.def.GF}
\end{equation}
where
\begin{subequations} \label{all}
\begin{align}
G^{\m{R}}_{jk}(t,t')
&=
-i\Theta(t-t')\braket{[c_j(t), c^\dagger_k(t')]_+}
\notag\\
&=
\big[G^{\m{A}}_{kj}(t',t)\big]^*
\label{eq.def.Gr}
,\\
G^<_{jk}(t,t') 
&= i\braket{c^\dagger_k(t') c_j(t)}
\label{eq.def.G<}
,\\
G^>_{jk}(t,t') 
&= -i\braket{c_j(t) c^\dagger_k(t')},
\label{eq.def.G>}
\end{align}
\end{subequations}
with $[A, B]_\pm = AB \pm BA$. In Eq.~\eqref{eq.def.GF}, $\bm{G}^{\m{R}}$, $\bm{G}^{\m{A}}$, $\bm{G}^<$, and $\bm{G}^>$ are, respectively, the retarded, advanced, lesser, and greater Green’s functions.

When the system is in a NESS, these nonequilibrium Green's functions satisfy the Dyson-Keldysh equations~\cite{RammerBook, StefanucciBook, HaugBook, CamsariBook},
\begin{align}
&
\bm{G}^{\m{R}(\m{A})}(\omega) =
\bm{G}^{\m{R}(\m{A})}_0(\omega)	+
\bm{G}^{\m{R}(\m{A})}_0(\omega)
\bm{\Sigma}^{\m{R}(\m{A})}(\omega)
\bm{G}^{\m{R}(\m{A})}(\omega)
\label{eq.Dyson.r}
,\\	
&
\bm{G}^{\lessgtr}(\omega) =
\bm{G}^{\m{R}}(\omega)
\bm{\Sigma}^{\lessgtr}(\omega)
\bm{G}^{\m{A}}(\omega).
\label{eq.Dyson.<}
\end{align}
Here, $\bm{G}^{\m{R}(\m{A})}_0$ denotes the bare Green's function of the {\it isolated} nanowire without electron scattering, given by
\begin{equation}
\bm{G}^{\m{R}(\m{A})}_0(\omega) = \frac{1}{[\omega \pm i\delta] \bm{1} -\bm{\m{H}}_0},
\end{equation}
where $\delta$ represents an infinitesimally small positive number, $\bm{1}$ denotes the $N\times N$ unit matrix, and $\bm{\m{H}}_0$ is the matrix representation of the Hamiltonian $H_0$ in Eq.~\eqref{eq.H0}. In Eqs.~\eqref{eq.Dyson.r} and \eqref{eq.Dyson.<}, $\bm{\Sigma}^{{\rm X}}$ is the $N\times N$ matrix self-energy correction, which consists of three parts,
\begin{equation}
\bm{\Sigma}^{\rm X}(\omega) =
\bm{\Sigma}^{\rm X}_{\rm lead}(\omega) +
\bm{\Sigma}^{\rm X}_{\rm imp}(\omega) +
\bm{\Sigma}^{\rm X}_{\rm ph}(\omega).
\end{equation}
Here, $\bm{\Sigma}^{\rm X}_{\rm lead}$, $\bm{\Sigma}^{\rm X}_{\rm imp}$, and $\bm{\Sigma}^{\rm X}_{\rm ph}$ describe the effects of reservoir couplings, elastic scattering from impurities, and inelastic scattering from phonons, respectively.

In the second-order Born approximation with respect to the tunneling amplitude $t_{\alpha}=t_{\rm lead}$, $\bm{\Sigma}^{\rm X}_{\rm lead}$ describing the couplings with the reservoirs takes the form~\cite{StefanucciBook, HaugBook, CamsariBook}
\begin{equation}
\Sigma^{\rm X}_{{\rm lead},jk}(\omega)
=
|t_{\rm lead}|^2\sum_{\bm{k}} 
\big[
\mathscr{G}^{\rm X}_{{\rm L},\bm{k}}(\omega) \delta_{j,1}+
\mathscr{G}^{\rm X}_{{\rm R},\bm{k}}(\omega) \delta_{j,N}
\big]
\delta_{j,k}.
\label{eq.self.lead.1}
\end{equation}
Here, the noninteracting Green's functions in the $\alpha$ ($={\rm L}, {\rm R}$) reservoir are given by~\cite{RammerBook, StefanucciBook, HaugBook, CamsariBook}
\begin{subequations} \label{all}
\begin{align}
&
\mathscr{G}^{\m{R}(\m{A})}_{\alpha,\bm{k}}(\omega) =
\frac{1}{\omega \pm i\delta -\xi_{\alpha,\bm{k}}}
\label{eq.nonint.G.RA}
,\\
&
\mathscr{G}^<_{\alpha,\bm{k}}(\omega) = 2\pi i \delta(\omega -\xi_{\alpha, \bm{k}}) f(\omega -\mu_\alpha)
\label{eq.nonint.G.<}
,\\
&
\mathscr{G}^>_{\alpha,\bm{k}}(\omega) = -2\pi i \delta(\omega -\xi_{\alpha, \bm{k}})
f(-\omega +\mu_\alpha).
\end{align}
\end{subequations}
Under the wide-band limit approximation~\cite{StefanucciBook, HaugBook, CamsariBook}, which assumes a constant density of states $\nu_\alpha(\omega)\equiv \nu$ in the reservoirs around the Fermi level $\omega=0$, the $\bm{k}$ summation in Eq.~\eqref{eq.self.lead.1} yields
\begin{subequations} \label{all}
\begin{align}
&
\bm{\Sigma}^{\m{R}(\m{A})}_{\alpha, {\rm lead}}(\omega)= \mp i \bm{\Gamma}_\alpha
\label{eq.self.lead.RA}
,\\
&
\bm{\Sigma}^<_{\alpha, {\rm lead}}(\omega)= 2i \bm{\Gamma}_\alpha f(\omega -\mu_\alpha)
\label{eq.self.lead.<}
,\\
&
\bm{\Sigma}^>_{\alpha, {\rm lead}}(\omega)= -2i \bm{\Gamma}_\alpha f(-\omega +\mu_\alpha),
\end{align}
\end{subequations}
with
\begin{align}
&
\bm{\Sigma}^{\rm X}_{{\rm lead}}(\omega)=
\sum_{\alpha={\rm L}, {\rm R}}\bm{\Sigma}^{\rm X}_{\alpha, {\rm lead}}(\omega)
,\\
&
\big[\bm{\Gamma}_{\rm L}\big]_{ij}=
\gamma_{\rm lead}\delta_{i,j}\delta_{i,1}
,\\
&
\big[\bm{\Gamma}_{\m{R}}\big]_{ij}=
\gamma_{\rm lead}\delta_{i,j}\delta_{i,N}
,\\
&
\gamma_{\rm lead} = \pi \nu |t_{\rm lead}|^2.
\label{eq.gam.lead}
\end{align}
This wide-band limit approximation is valid when the energy dependence of the density of states $\nu_\alpha(\omega)$ in the reservoirs can be ignored around the Fermi level $\omega=0$, within the range of applied bias voltage $eV$~\cite{Kawamura2020}. This condition is well satisfied under typical experimental conditions, where the reservoirs are made of normal metals such as gold or copper, which exhibit a smooth and broad density of states near the Fermi level, and the applied bias voltages are on the order of a few millivolts at low temperatures.

We deal with the self-energy correction $\bm{\Sigma}^{\rm X}_{\rm imp}$ describing  electron-impurity scattering effects within the self-consistent Born approximation~\cite{RammerBook, HaugBook, CamsariBook}, which yields
\begin{subequations} \label{all}
\begin{align}
\Sigma^{\rm X}_{{\rm imp},jk}(\omega)
&=
U_{{\rm imp},j}\, U_{{\rm imp},k}\, G^{\rm X}_{jk}(\omega)
\\
&=
\delta U_{{\rm imp},j}\, \delta U_{{\rm imp},k}\, 
G^{\rm X}_{jk}(\omega).
\label{eq.self.imp.1}
\end{align}
\end{subequations}
In obtaining Eq.~\eqref{eq.self.imp.1}, we have used Eq.~\eqref{eq.Uimp} and $\bar{U}_{\rm imp}=0$. We then perform the random average over impurity configurations.
Assuming no spatial correlations between impurities, the binary impurity variables satisfy
\begin{equation}
\braket{n_{{\rm imp},j}\, n_{{\rm imp},k}}_{\rm imp} =
\begin{cases}
\bar{n}_{\rm imp} \hspace{0.3cm}(j=k)\\
\bar{n}^2_{\rm imp} \hspace{0.3cm}(j\neq k)
\end{cases},
\end{equation}
which leads to
\begin{align}
\braket{\delta U_{{\rm imp},j}\, \delta U_{{\rm imp},k}}_{\rm imp}
&=
u^2_{\rm imp}\, \big[\braket{n_{{\rm imp},j}\, n_{{\rm imp},k}}_{\rm imp} -\bar{n}^2_{\rm imp} \big]
\notag\\
&=
u^2_{\rm imp}\, \bar{n}_{\rm imp} [1-\bar{n}_{\rm imp}]\, \delta_{j,k}
\notag\\
&\equiv
\gamma^2_{\rm imp}\, \delta_{j,k}.
\label{eq.imp.ave}
\end{align}
Here, $\braket{\cdot}_{\rm imp}$ denotes the average over different impurity configurations. Applying Eq.~\eqref{eq.imp.ave} to Eq.~\eqref{eq.self.imp.1}, we obtain~\cite{Alessandro2008}
\begin{align}
\bm{\Sigma}^{\rm X}_{\rm imp}(\omega)
&=
\gamma^2_{\rm imp}
\begin{pmatrix}
G^{\rm X}_{11}(\omega) && 0 \\
& \ddots & \\
0 && G^{\rm X}_{NN}(\omega)
\end{pmatrix}
\notag\\
&=
\gamma^2_{\rm imp}
\bm{1} \circ \bm{G}^{\rm X}(\omega).
\label{eq.self.imp.2}
\end{align}
Here, ``$\circ$" denotes the Hadamard (element-wise) product. This operation extracts the diagonal part of $\bm{G}^{\rm X}(\omega)$ by zeroing out all off-diagonal elements. The parameter $\gamma^2_{\rm imp}= u^2_{\rm imp}\, \bar{n}_{\rm imp} [1-\bar{n}_{\rm imp}]$ characterizes the impurity scattering strength: $\gamma_{\rm imp}=0$ (large $\gamma_{\rm imp}$) corresponds to the ballistic (diffusive) limit. We note that in the presence of strong impurity scattering, the system is characterized by a rapidly varying potential, leading to electron localization within potential walls (Anderson localization)~\cite{heikkilaBook, Anderson1958}. However, this localized regime lies beyond the scope of this paper.

Within the self-consistent Born approximation~\cite{RammerBook, HaugBook}, the self-energy $\bm{\Sigma}^{\rm X}_{\rm ph}$  describing electron-phonon scattering effects takes the form
\begin{subequations} \label{all}
\begin{align}
&
\bm{\Sigma}^{\m{R}(\m{A})}_{\rm ph}(\omega)=ig_{\rm ph}^2\int_{-\infty}^\infty\frac{d\nu}{2\pi}\Big[
D^{\m{R}(\m{A})}(\nu) \bm{1}\circ \bm{G}^<(\omega-\nu)
\notag\\
&\hspace{2cm}
+
D^{\m{R}(\m{A})}(\nu) \bm{1}\circ \bm{G}^{\m{R}(\m{A})}(\omega-\nu) 
\notag\\
&\hspace{2cm}
+
D^<(\nu) \bm{1}\circ \bm{G}^{\m{R}(\m{A})}(\omega-\nu)
\Big]
\label{eq.self.ph.r}
,\\
&
\bm{\Sigma}^\lessgtr_{\rm ph}(\omega)= ig_{\rm ph}^2\int_{-\infty}^\infty\frac{d\nu}{2\pi} D^\lessgtr(\nu) \bm{1}\circ \bm{G}^\lessgtr(\omega-\nu).
\label{eq.self.ph.<}
\end{align}
\end{subequations}
Here, $D^{\rm X}$ is the phonon Green's function, given by~\cite{RammerBook, HaugBook}
\begin{subequations} \label{all}
\begin{align}
&
D^{\m{R}(\m{A})}(\nu)= \frac{1}{\nu-\Omega_0 \pm i\delta} +\frac{1}{\nu+\Omega_0 \pm i\delta}
,\\
&
D^\lessgtr(\nu) = -2\pi i  \Big[\big[n_{\rm B}(\Omega_0)+1\big]
\delta(\nu \pm \Omega_0) +n_{\rm B}(\Omega_0)\delta(\nu \mp \Omega_0) \Big],
\label{eq.Dph.<}
\end{align}
\end{subequations}
with the Bose-Einstein distribution function,
\begin{equation}
n_{\rm B}(\nu) = \frac{1}{e^{\nu/T_{\rm env}}-1}.
\end{equation}
In deriving $\bm{\Sigma}^{\rm X}_{\rm ph}$, we have assumed that phonons are unperturbed by electron-phonon couplings and maintain thermal equilibrium at temperature $T_{\rm env}$. This is a simplification, but it can be physically justified when the nanowire is on a substrate, where phonons in the wire strongly couple to those in the substrate and remain equilibrated~\cite{heikkilaBook}. Substituting Eq.~\eqref{eq.Dph.<} into Eq.~\eqref{eq.self.ph.<} yields
\begin{align}
\bm{\Sigma}^\lessgtr_{\rm ph}(\omega) &= 
g_{\rm ph}^2\Big[\big[n_{\rm B}(\Omega_0)+1\big] \bm{1}\circ \bm{G}^\lessgtr(\omega \pm \Omega_0)
\notag\\
&\hspace{2cm} 
+n_{\rm B}(\Omega_0) \bm{1}\circ \bm{G}^\lessgtr(\omega \mp \Omega_0)\Big].
\label{eq.self.ph.<.2}
\end{align}
We note that unlike the lesser and greater components $\bm{\Sigma}^\lessgtr_{\rm ph}$, the $\nu$ integral in Eq.~\eqref{eq.self.ph.r} cannot be performed analytically. The efficient numerical computation of the retarded component $\bm{\Sigma}^{\m{R}}_{\rm ph}$ is detailed in Appendix~\ref{sec.app.KK}.

The dressed Green's functions are obtained by incorporating all self-energy corrections into the Dyson equations~\eqref{eq.Dyson.r} and \eqref{eq.Dyson.<}. From Eq.~\eqref{eq.Dyson.r}, we have
\begin{align}
\bm{G}^{\m{R}}(\omega) 
&=
\frac{1}{\omega \bm{1} -\bm{\m{H}}_0 -\bm{\Sigma}^{\m{R}}_{\rm lead}(\omega)-\bm{\Sigma}^{\m{R}}_{\rm imp}(\omega)-\bm{\Sigma}^{\m{R}}_{\rm ph}(\omega)}
\notag\\
&\equiv \bm{T}^{-1}.
\label{eq.Gr.full}
\end{align}
Since the bare Hamiltonian $\bm{\m{H}}_0$ is tridiagonal and all self-energy terms are diagonal matrices, $\bm{T}$ retains a tridiagonal structure. This tridiagonal structure of $\bm{T}$ allows for efficient and stable computation of $\bm{T}^{-1}$. The numerical implementation is presented in Appendix~\ref{sec.app.Inv.Tri}.

The lesser Green's function $\bm{G}^<$ is obtained by substituting the dressed retarded Green's function in Eq.~\eqref{eq.Gr.full} into the Dyson equation~\eqref{eq.Dyson.<}. Noting that $\bm{\Sigma}^<=\bm{\Sigma}^<_{\rm lead}+\bm{\Sigma}^<_{\rm imp}+\bm{\Sigma}^<_{\rm ph}$ is a diagonal matrix, we have
\begin{subequations} \label{all}
\begin{align}
G^<_{jj}(\omega)
&=
\sum_{l,m=1}^N 
G^{\m{R}}_{jl}(\omega) \Sigma^<_{lm}(\omega)
G^{\m{A}}_{mj}(\omega)
\\ 
&=
\sum_{l=1}^N |G^{\m{R}}_{jl}(\omega)|^2 \Sigma^<_{ll}(\omega).
\label{eq.G<.jk}
\end{align}
\end{subequations}
Since the self-energy corrections $\bm{\Sigma}^{\rm X}_{\rm imp}$ and $\bm{\Sigma}^{\rm X}_{\rm ph}$ involve the dressed Green's function $\bm{G}^{\rm X}$, a self-consistent calculation is required. To accelerate the convergence of this self-consistent loop, we employ the restarted Pulay mixing scheme~\cite{Pulay1980, Phanisri2015, Amartya2016}. For a detailed description of this technique, see Appendix~\ref{sec.app.Pulay}.

Once we obtain the dressed Green's functions, the local density of states $\nu_j(\omega)$, the filling fraction $n_j$, and  the nonequilibrium distribution function $f^{\rm neq}_j(\omega)$ at site $j$ are, respectively, obtained as~\cite{RammerBook, HaugBook, StefanucciBook, CamsariBook, Ness2013, Ness2014}
\begin{align}
& \nu_j(\omega) = -\frac{1}{\pi} {\rm Im} G^{\m{R}}_{jj}(\omega)
\label{eq.LDOS}
,\\[4pt]
& n_j = -i\int_{-\infty}^\infty \frac{d\omega}{2\pi}G^<_{jj}(\omega)
,\\[4pt]
& f^{\rm neq}_j(\omega) = \frac{-iG^<_{jj}(\omega)}{2\pi \nu_j(\omega)}
\label{eq.fneq}.
\end{align}
We note that the charge current through the wire can also be evaluated using the dressed Green's function $\bm{G}^{\rm X}$, as detailed in Appendix~\ref{sec.app.transport}.

The requirement of charge neutrality in the wire imposes the condition~\cite{heikkilaBook}
\begin{equation}
\Delta n_j = n_j -n_j^0 =0, \hspace{0.5cm}(j=1,\cdots,N)
\label{eq.null.charge}
\end{equation}
where $n_j^0$ represents the filling fraction in the absence of bias voltage $V$ (that is, $\mu_{\rm L}=\mu_{\rm R}$). This local charge neutrality condition, widely used in the literature~\cite{heikkilaBook, KopninBook}, effectively approximates the result of solving the Poisson equation self-consistently, and is justified when the screening length is sufficiently short compared to the system size. Because the electrostatic potential $\varphi_j$ enters the Hamiltonian in Eq.~\eqref{eq.H0} and thus affects the Green's functions, Eq.~\eqref{eq.null.charge} must be solved self-consistently with respect to $\varphi_j$. As explained in Appendix~\ref{sec.app.workflow}, we determine $\varphi_j$ by solving the nonlinear self-consistent equation~\eqref{eq.null.charge} using the Broyden method~\cite{Broyden1965, Broyden1967}.

%%%%%%%%%%%%%%%%%%%%%%%%%%%%%%%%
\begin{figure*}[t]
\centering
\includegraphics[width=15.5cm]{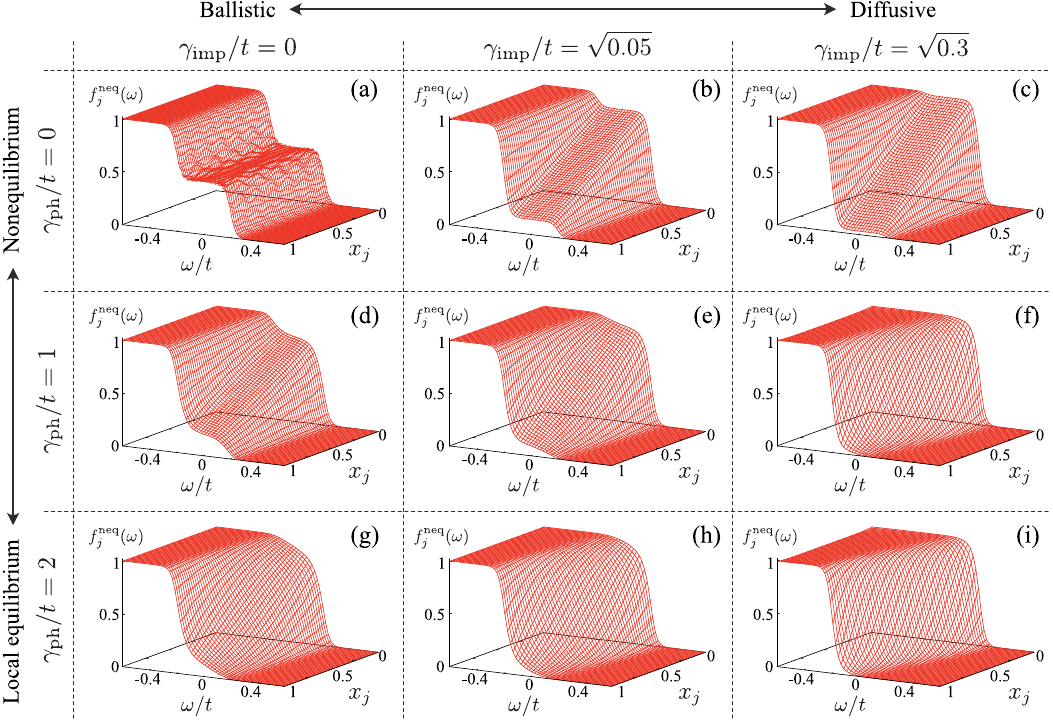}
\caption{Calculated electron distribution function $f_j^{\rm neq}(\omega)$ in a nanowire under bias voltage. The position $x_j$ along the wire is defined by Eq.~\eqref{eq.xj}. We show results for different values of impurity scattering strength $\gamma_{\rm imp}$ and electron-phonon coupling strength $\gamma_{\rm ph}$. We set $N=201$, $eV/t = 0.4$, $\gamma_{\rm lead}/t=1$, $\Omega_0/t=0.05$ and $T_{\rm env}/t=0.02$. These values are used in the following figures unless otherwise mentioned.}
\label{fig.fneq}
\end{figure*}
%%%%%%%%%%%%%%%%%%%%%%%%%%%%%%%%

%%%%%%%%%%%%%%%%%%%%%%%%%%%%%%%%%%%%%%%%%%%%%%%
\section{Nonequilibrium Distribution Function in a Voltage-Biased Wire \label{sec.result}}

Figure~\ref{fig.fneq} shows the calculated electron distribution function $f_j^{\rm neq}(\omega)$ in a wire under bias voltage. In this figure, we set $eV/t = 0.4$,  corresponding to the linear transport regime (see Appendix~\ref{sec.app.transport}). In the following, we discuss the effects of elastic and inelastic electron scattering on the distribution function in turn.

%%%%%%%%%%%%%%%%%%%%%%%%%%%%%%%%%%%%%%%%%%%%%%%%%%%%%%%%%%%%%%%%
\subsection{Crossover from the Ballistic to the Diffusive Regime \label{sec.result.imp}}

We first discuss the changes in the form of the distribution function $f_j^{\rm neq}(\omega)$ due to elastic scattering from impurities, shown in Fig.~\ref{fig.fneq}(a)-(c). In the ballistic limit ($\gamma_{\rm ph} = \gamma_{\rm imp}=0$), where electrons traverse the wire without any scattering, the local density of states $\nu_j(\omega)$ in Eq.~\eqref{eq.LDOS} can be expressed as
\begin{align}
\nu_j(\omega)
&=
\frac{i}{2\pi}\big[G^{\m{R}}_{jj}(\omega) -G^{\m{A}}_{jj}(\omega)\big]
\notag\\
&=	
\frac{i}{2\pi} \sum_{l,m=1}^N G^{\m{R}}_{jl}(\omega)\big[\Sigma^{\m{R}}_{{\rm lead}, lm} -\Sigma^{\m{A}}_{{\rm lead}, lm}\big](\omega) G^{\m{A}}_{mj}(\omega)
\notag\\
&=	
\frac{\gamma_{\rm lead}}{\pi}
\big[|G^{\m{R}}_{j1}(\omega)|^2 +|G^{\m{R}}_{jN}(\omega)|^2\big].
\label{eq.LDOS.ballistic}
\end{align}
Here, we have used~\cite{HaugBook}
\begin{align}
& G^{\m{R}}_{jk}(\omega) = \big[G^{\m{A}}_{kj}(\omega)\big]^*
,\\
& G^{\m{R}}_{jj}(\omega) -G^{\m{A}}_{jj}(\omega) =
\sum_{l,m=1}^N G^{\m{R}}_{jl}(\omega)\big[\Sigma^{\m{R}}_{lm} -\Sigma^{\m{A}}_{lm}\big](\omega) G^{\m{A}}_{mj}(\omega).
\label{eq.formula.GR.GA}
\end{align}
In the ballistic limit, the lesser Green's function $G^<_{jj}(\omega)$ in Eq.~\eqref{eq.G<.jk} is reduced to
\begin{align}
G^<_{jj}(\omega) &= 2i\gamma_{\rm lead}
\Big[f(\omega-\mu_{\rm L})|G^{\m{R}}_{j1}(\omega)|^2
\notag\\
&\hspace{2.8cm}
+ f(\omega-\mu_{\rm R})|G^{\m{R}}_{jN}(\omega)|^2 \Big].
\label{eq.G<.jj.ballistic}
\end{align}
Using Eqs.~\eqref{eq.fneq}, \eqref{eq.LDOS.ballistic} and \eqref{eq.G<.jj.ballistic}, we obtain the nonequilibrium distribution function $f^{\rm neq}_j(\omega)$ in the ballistic limit as
\begin{equation}
f_j^{\rm neq}(\omega)= w_j(\omega)f(\omega -\mu_{\rm L}) + \big[1-w_j(\omega)\big]f(\omega-\mu_{\rm R}),
\label{eq.fneq.ballistic}
\end{equation}
where we define the weight function as
\begin{equation}
w_j(\omega)=\frac{|G^{\m{R}}_{j1}(\omega)|^2}{|G^{\m{R}}_{j1}(\omega)|^2 +|G^{\m{R}}_{jN}(\omega)|^2}.
\label{eq.wj}
\end{equation}
Equation~\eqref{eq.fneq.ballistic} clearly shows that the distribution function $f_j^{\rm neq}(\omega)$, which describes the probability of observing an electron with energy $\omega$ at site $j$, is given by the sum of the two probabilities: (1) $w_j(\omega) f(\omega -\mu_{\rm L})$, the probability of an electron with energy $\omega$ propagating from the left reservoir, and (2) $[1-w_j(\omega)] f(\omega -\mu_{\rm R})$, the probability of an electron propagating from the right reservoir.

\begin{figure}[t]
\centering
\includegraphics[width=8.2cm]{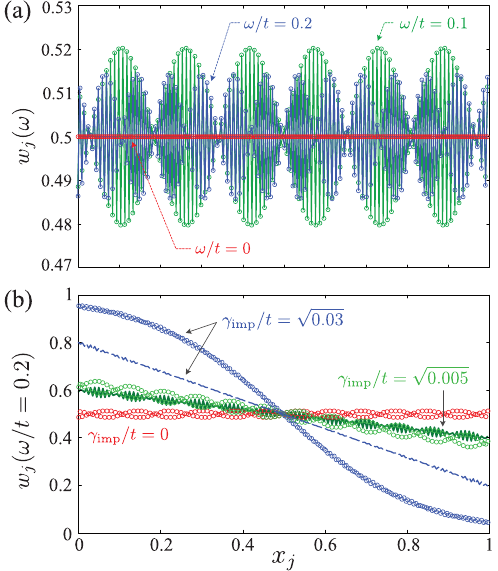}
\caption{(a) Position $x_j$ dependence of the weight function $w_j(\omega)$ in Eq.~\eqref{eq.wj}. We show the results for $\omega/t=0$, $0.1$, and $0.2$ in the ballistic limit ($\gamma_{\rm imp}=0$). (b) Impurity scattering strength $\gamma_{\rm imp}$ dependence of the weight function $w_j(\omega)$. As a typical example, we take $\omega/t=0.2$. Circles show $w_j$ calculated directly from the definition in Eq.~\eqref{eq.wj}. For comparison, solid and dashed lines show $w_j$ obtained by fitting the calculated distribution function $f^{\rm neq}_j(\omega)$ using Eq.~\eqref{eq.fneq.ballistic} for $\gamma_{\rm imp}/t=\sqrt{0.005}$ and $\sqrt{0.03}$, respectively. As $\gamma_{\rm imp}$ increases, the deviation between the two becomes more pronounced, reflecting the breakdown of  Eq.~\eqref{eq.fneq.ballistic} in the diffusive regime.}
\label{fig.wj}
\end{figure}

In the ballistic limit, the amplitude $|G^{\m{R}}_{j1}(\omega)|^2$, which represents the propagation probability of electron with energy $\omega$ from site 1 to $j$, should be independent of the site index $j$ due to the absence of scattering. Therefore, we expect $|G^{\m{R}}_{j1}(\omega)|^2 \simeq |G^{\m{R}}_{jN}(\omega)|^2 \big(\simeq |G^{\m{R}}_{N1}(\omega)|^2\big)$, leading to $w_j(\omega) \simeq 0.5$. This is verified in Fig.~\ref{fig.wj}(a), which shows that $w_j(\omega)$ is constant over space, except for the minor oscillations around $w_j(\omega)=0.5$. Using this fact, one can approximate Eq.~\eqref{eq.fneq.ballistic} as
\begin{equation}
f^{\rm neq}_j(\omega) \simeq \frac{1}{2}\big[f(\omega -\mu_{\rm L}) +f(\omega -\mu_{\rm R}) \big].
\label{eq.fneq.ballistic.app}
\end{equation}
Thus, in the ballistic limit, the distribution function $f^{\rm neq}_j(\omega)$ is simply expressed as the average of the Fermi-Dirac distribution functions in the left and right reservoirs, and does not depend on the position $x_j$ along the wire~\cite{heikkilaBook, GueronThesis, HuardThesis, AnneThesis, PierreThesis}, as shown in Fig.~\ref{fig.fneq}(a). We also see from Eq.~\eqref{eq.fneq.ballistic.app} that the distribution function $f_j^{\rm neq}(\omega)$ exhibits the two-step structure at low temperatures ($T_{\rm env} \ll \mu_{\rm L}-\mu_{\rm R}$), reflecting the different Fermi-Dirac distribution function $f(\omega -\mu_\alpha)$ in the $\alpha$ reservoir.

%%%%%%%%%%%%%%%%%%%%%%%%%%%%%%%%
%%%%%%%%%%%%%%%%%%%%%%%%%%%%%%%%

We briefly note that the spatial oscillations in the weight function $w_j(\omega)$ shown in Fig.~\ref{fig.wj}(a) arise from Fabry-Perot-like interference between the left and right reservoirs, which act as potential barriers~\cite{ScheerBook, StegmannThesis}. These oscillations in the weight function $w_j(\omega)$ result in the oscillations in the distribution function $f^{\rm neq}_j(\omega)$ around $|\omega| \lesssim [\mu_{\rm L}-\mu_{\rm R}]/2=0.2 t$, as shown in Fig.~\ref{fig.fneq}(a).

%%%%%%%%%%%%%%%%%%%%%%%%%%%%%%%%
\begin{figure}[t]
\centering
\includegraphics[width=8.2cm]{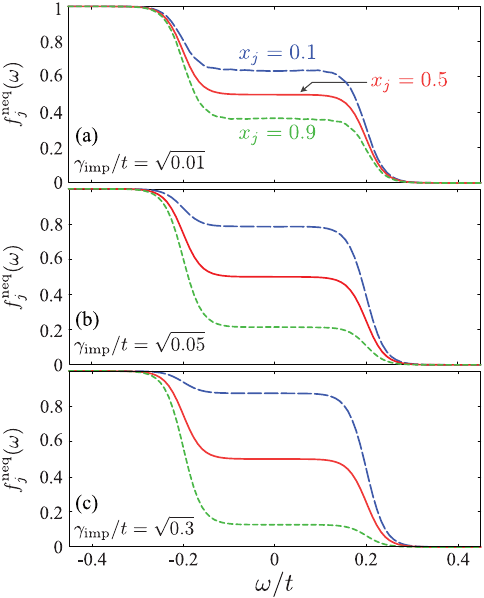}
\caption{Calculated distribution function $f_j^{\rm neq}(\omega)$ for (a) $\gamma_{\rm imp}/t=\sqrt{0.01}$, (b) $\gamma_{\rm imp}/t=\sqrt{0.05}$, and (c) $\gamma_{\rm imp}/t=\sqrt{0.3}$. We show the distribution function at $x_j=0.1$ (near the left reservoir), $x_j=0.5$ (in the middle of the wire), and $x_j=0.9$ (near the right reservoir). We set $\gamma_{\rm ph}=0$ for all panels.}
\label{fig.imp.x}
\end{figure}
%%%%%%%%%%%%%%%%%%%%%%%%%%%%%%%%

Figures~\ref{fig.fneq}(b) and (c) show that elastic scattering from impurities results in a spatially varying distribution function $f^{\rm neq}_j(\omega)$. Figure~\ref{fig.imp.x} shows the distribution function at three positions: $x_j=0.1$ (near the left reservoir), $x_j=0.5$ (in the middle of the wire), and $x_j=0.9$ (near the right reservoir), for different impurity scattering strengths $\gamma_{\rm imp}$. In the presence of electron-impurity scattering, electrons traverse the wire via a random walk process. As a result, the distribution function at site $j$ more strongly reflects the Fermi-Dirac distribution function in the reservoir closer to site $j$.

\begin{figure}[t]
\centering
\includegraphics[width=8.2cm]{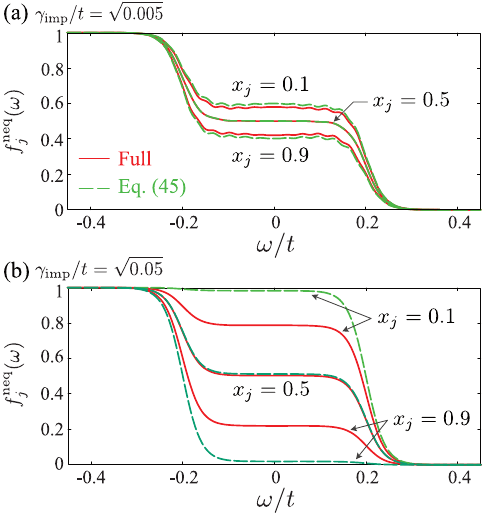}
\caption{Comparison of the distribution function $f^{\rm neq}_j(\omega)$ at three positions ($x_j=0.1$, $0.5$, and $0.9$) obtained from the full self-consistent calculation (solid line) and the approximate expression in Eq.~\eqref{eq.fneq.ballistic} (dashed line). We set (a) $\gamma_{\rm imp}/t=\sqrt{0.005}$ and (b) $\gamma_{\rm imp}/t = \sqrt{0.05}$.
}
\label{fig.app.fw}
\end{figure}

Although the distribution function $f^{\rm neq}_j(\omega)$ deviates from Eq.~\eqref{eq.fneq.ballistic} in the presence of impurity scattering, its overall behavior can still be reasonably described by Eq.~\eqref{eq.fneq.ballistic} for sufficiently small $\gamma_{\rm imp}$, as demonstrated in Fig.~\ref{fig.app.fw}(a). This allows us to analyze the spatial dependence of the distribution function $f^{\rm neq}_j(\omega)$ in the ballistic-diffusive crossover regime using the weight function $w_j(\omega)$ in Eq.~\eqref{eq.wj}. Figure~\ref{fig.wj}(b) shows that in the presence of impurity scattering ($\gamma_{\rm imp}\neq 0$), the weight function $w_j(\omega)$ decreases with increasing $x_j$. Since the weight function $w_j(\omega)$ physically represents the probability of an electron with energy $\omega$ propagating from the left reservoir to site $j$, Fig.~\ref{fig.wj} (b) indicates that information about the distribution function $f(\omega -\mu_\alpha)$ in the $\alpha$ reservoir is gradually lost due to elastic scattering from impurities as electrons propagate away from the $\alpha$ reservoir. 

For larger impurity scattering strength $\gamma_{\rm imp}$, the approximate expression in Eq.~\eqref{eq.fneq.ballistic} deviates from the full self-consistent calculation, as shown in Fig.~\ref{fig.app.fw}(b). A similar trend is observed in Fig.~\ref{fig.wj}(b), where the weight function $w_j(\omega)$  obtained by fitting the self-consistently calculated distribution function with Eq.~\eqref{eq.fneq.ballistic} increasingly deviates from the one directly computed from Eq.~\eqref{eq.wj} as $\gamma_{\rm imp}$ increases.

This discrepancy arises because the approximate expression in Eq.~\eqref{eq.fneq.ballistic} retains only the leading-order contribution of impurity scattering. When $\gamma_{\rm imp}\neq 0$, the diagonal component $G^<_{jj}(\omega)$ of Eq.~\eqref{eq.G<.jk} is expanded as 
\begin{subequations} \label{all}
\begin{align}
G^<_{jj}(\omega)
&=
Q_{jj}(\omega)	
+
\sum_{k=1}^N|G^{\m{R}}_{jk}(\omega)|^2 \Sigma^<_{{\rm imp},kk}(\omega)
\\
&=
Q_{jj}(\omega)	
+
\gamma_{\rm imp}\sum_{k=1}^N|G^{\m{R}}_{jk}(\omega)|^2 Q_{kk}(\omega)
\notag\\
&\hspace{0.2cm}
+
\gamma^2_{\rm imp}
\sum_{k_1=1}^N\sum_{k_2=1}^N
|G^{\m{R}}_{jk_1}(\omega)|^2 
|G^{\m{R}}_{k_1k_2}(\omega)|^2 
Q_{k_2k_2}(\omega)
\notag\\
&\hspace{0.2cm}
+
\cdots,
\label{eq.G<.expandion}
\end{align}
\end{subequations}
with
\begin{subequations} \label{all}
\begin{align}
Q_{jj}(\omega)
&=
\big[
\bm{G}^{\m{R}}(\omega)
\bm{\Sigma}^{<}_{\rm lead}(\omega)
\bm{G}^{\m{A}}(\omega)
\big]_{jj}
\\
&=
2i\gamma_{\rm lead}
\Big[f(\omega-\mu_{\rm L})|G^{\m{R}}_{j1}(\omega)|^2
\notag\\
&\hspace{2.5cm}
+ f(\omega-\mu_{\rm R})|G^{\m{R}}_{jN}(\omega)|^2 \Big].
\label{eq.Qjj}	
\end{align}
\end{subequations}
The first term $Q_{jj}(\omega)$ in Eq.~\eqref{eq.G<.expandion} represents the contribution from electrons propagating directly from the left (right) lead to site $j$, without experiencing impurity scattering. The approximate expression in Eq.~\eqref{eq.fneq.ballistic} captures only this leading-order term, as is evident by comparing Eq.~\eqref{eq.Qjj} with Eq.~\eqref{eq.G<.jj.ballistic}. 

In contrast, the second term in Eq.~\eqref{eq.G<.expandion} corresponds to processes in which an electron propagating from the left (right) lead to site $k$ with amplitude $|G^{\m{R}}_{k,1(N)}(\omega)|^2$ is scattered by an impurity at site $k$ with amplitude $\gamma_{\rm imp}$, and then propagates to site $j$ with amplitude $|G^{\m{R}}_{j,k}(\omega)|$. Similarly, the third term describes contributions from electrons that are scattered twice by impurities at sites $k_1$ and $k_2$ before reaching site $j$. As $\gamma_{\rm imp}$ increases, these higher-order scattering processes become increasingly important, leading to the discrepancy observed in Fig.~\ref{fig.app.fw}(b), as well as Fig.~\ref{fig.wj}(b).

%%%%%%%%%%%%%%%%%%%%%%%%%%%%%%%%
%%%%%%%%%%%%%%%%%%%%%%%%%%%%%%%%

%%%%%%%%%%%%%%%%%%%%%%%%%%%%%%%%
\begin{figure}[t]
\centering
\includegraphics[width=8.2cm]{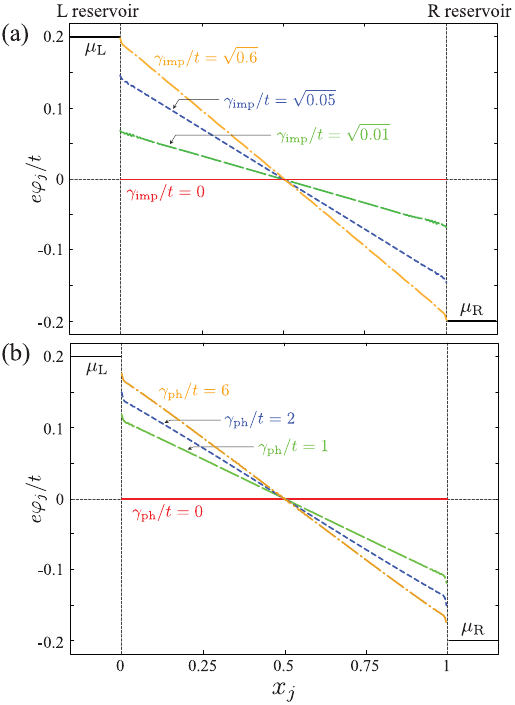}
\caption{Profile of the electrostatic potential $\varphi_j$ for different values of (a) impurity scattering strength $\gamma_{\rm imp}$ and (b) electron-phonon coupling strength $\gamma_{\rm ph}$. The electrochemical potential in the $\alpha$ reservoir is fixed at $\mu_\alpha/t = \pm 0.2$. In panel (a), we set $\gamma_{\rm ph}=0$, while in panel (b), we set $\gamma_{\rm imp}=0$.}
\label{fig.phi}
\end{figure}
%%%%%%%%%%%%%%%%%%%%%%%%%%%%%%%%

%%%%%%%%%%%%%%%%%%%%%%%%%%%%%%%%
\begin{figure*}[t]
\centering
\includegraphics[width=15.5cm]{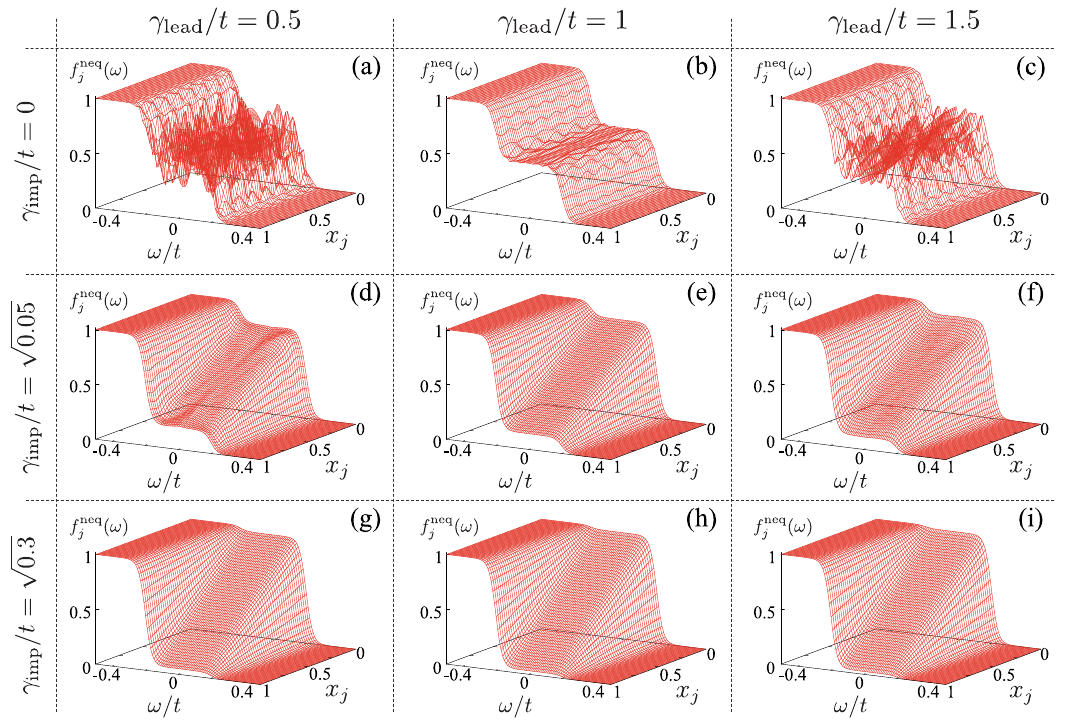}
\caption{Distribution function $f_j^{\rm neq}(\omega)$ for different values of the wire-reservoir coupling parameter $\gamma_{\rm lead}$ and impurity scattering strength $\gamma_{\rm imp}$. In all panels, we set $\gamma_{\rm ph}=0$. In the ballistic limit ($\gamma_{\rm imp}=0$), the distribution function exhibits pronounced oscillations that depend sensitively on $\gamma_{\rm lead}$, while in the diffusive regime these oscillations are suppressed and the distribution becomes nearly insensitive to $\gamma_{\rm lead}$. Note that panels (b), (e), and (h) correspond to the results shown in Figs.~\ref{fig.fneq}(a), (b), and (c), respectively.}
\label{fig.fneq.lead}
\end{figure*}
%%%%%%%%%%%%%%%%%%%%%%%%%%%%%%%%

We see from Fig.~\ref{fig.fneq}(c) that in the diffusive (sufficiently large $\gamma_{\rm imp}$) limit, the distribution function linearly interpolates between the Fermi-Dirac distribution functions in the left and right reservoirs at each energy $\omega$, which can be expressed as
\begin{equation}
f_j^{\rm neq}(\omega)=
\big[1-x_j\big] f(\omega -\mu_{\rm L})
+
x_jf(\omega-\mu_{\rm R}).
\label{eq.fneq.stairstep}
\end{equation}
This nonequilibrium distribution function is obtained by solving the transport equation~\eqref{eq.Boltzmann.diffusive} in the absence of elastic scattering~\cite{GueronThesis, Bronn2013},
\begin{equation}
D\partial_x^2 f^{\rm neq}_x(\omega) =0,
\label{eq.Boltzmann}
\end{equation}
under the boundary conditions in Eqs.~\eqref{eq.BC1} and \eqref{eq.BC2}. Thus, Figs.~\ref{fig.fneq}(a)-(c) demonstrate that the distribution functions in the ballistic-diffusive crossover regime are well described within our framework.

Figure~\ref{fig.phi}(a) shows the electrostatic potential $\varphi_j$ along the wire in the ballistic-diffusive crossover regime. In the ballistic limit ($\gamma_{\rm imp}=0$), the voltage drops at the contacts ($x_j=0$ and $x_j=1$) between the wire and the reservoirs, while the potential $\varphi_j$ is constant in the bulk of the wire~\cite{ImryBook, DattaBook, ScheerBook, StegmannThesis}. In this case, the resistance dominantly governed by contributions from the contacts, which is known as the contact resistance~\cite{StegmannThesis, DattaBook1995}. On the other hand, as the impurity scattering strength $\gamma_{\rm imp}$ increases, the potential $\varphi_j$ profile changes from the flat profile with large jumps at the contacts to a smooth linear profile connecting the electrochemical potentials $\mu_{\alpha={\rm L}, {\rm R}}$ in the reservoirs. In this regime, multiple elastic scattering events lead to diffusive Ohmic transport, characterized by a resistance proportional to the wire length. The crossover from ballistic to Ohmic transport can also be directly observed through electron transport properties, as discussed in Appendix~\ref{sec.app.transport}.

To clarify how the impurity-scattering parameter $\gamma_{\rm imp}$ controls this ballistic-diffusive crossover, it is useful to estimate the mean free path $l_{\rm imp}$ within our one-dimensional tight-binding model. The elastic scattering rate due to impurities is given by
\begin{equation}
\frac{1}{\tau_{\rm imp}(\omega)} = 2\pi \gamma_{\rm imp}^2 \nu(\omega),
\end{equation}
where the density of states $\nu(\omega)$ of the chain is 
\begin{equation}
\nu(\omega) = \frac{1}{\pi} \frac{1}{\sqrt{4t^2 -\omega^2}}\Theta(2t-|\omega|).
\end{equation}
Since the electrochemical potentials of the reservoirs lie near the band center, we may evaluate the relaxation time at $\omega=0$. At half filling ($k_{\rm F}=\pi/2$), we have
\begin{equation}
\frac{1}{\tau_{\rm imp}(0)} = \frac{\gamma_{\rm imp}^2}{t}.
\end{equation}
Using the Fermi velocity $v_{\rm F}=2t \sin k_{\rm F}=2t$, we can evaluate the mean free path as
\begin{equation}
l_{\rm imp} \sim v_{\rm F}\tau_{\rm imp}(0) \sim \frac{2}{(\gamma_{\rm imp}/t)^2}.
\end{equation}
Because the lattice constant is set to $a=1$, this expression directly yields the mean free path measured in units of lattice sites.

This estimate provides a clear interpretation of the ballistic–diffusive crossover. For example, in Fig.~\ref{fig.fneq}(b), $\gamma_{\rm imp}/t = \sqrt{0.05}$ gives $l_{\rm imp} \simeq 40$ sites. With the wire length of $L=201$ sites, an electron experiences only a few scattering events while traversing the wire, resulting in an intermediate regime between ballistic and diffusive transport. In contrast, $\gamma_{\rm imp}/t = \sqrt{0.3}$ in Fig.~\ref{fig.fneq}(c) gives $l_{\rm imp} \simeq 7$ sites, and $\gamma_{\rm imp}/t = \sqrt{0.6}$ in Fig.~\ref{fig.phi}(a) gives $l_{\rm imp} \simeq 3$ sites. In these cases, an electron undergoes tens of scattering events before reaching the opposite electrode, so that individual scattering processes are effectively averaged out and the transport becomes diffusive.

\begin{figure}[t]
\centering
\includegraphics[width=8.2cm]{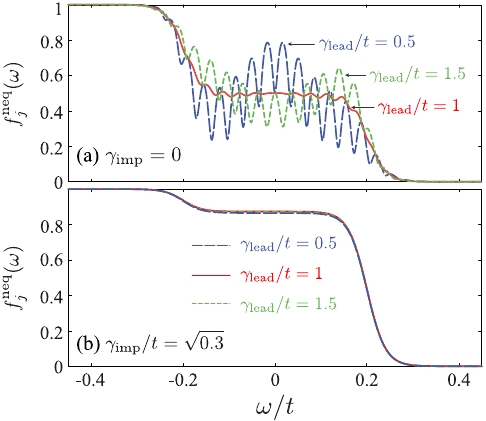}
\caption{Distribution function $f^{\rm neq}_j(\omega)$ at position $x_j=0.1$ for different values of the wire-reservoir coupling parameter $\gamma_{\rm lead}$, in (a) the ballistic limit ($\gamma_{\rm imp}=0$) and (b) the diffusive regime ($\gamma_{\rm imp}/t=\sqrt{0.3}$).
}
\label{fig.fw.lead.x01}
\end{figure}

Figure~\ref{fig.fneq.lead} shows how the distribution function $f^{\rm neq}_j(\omega)$ depends on $\gamma_{\rm lead}$ in Eq.~\eqref{eq.gam.lead} in both the ballistic and diffusive regimes. In the ballistic limit, as seen in Figs.~\ref{fig.fneq.lead}(a)–(c), the shape of the distribution function is highly sensitive to $\gamma_{\rm lead}$, while in the diffusive regime, as seen in Figs.~\ref{fig.fneq.lead}(g)–(i), it becomes much less sensitive. Figure~\ref{fig.fw.lead.x01}(a) shows this behavior more clearly by showing the distribution function at $x_j=0.1$ in the ballistic limit for several values of $\gamma_{\rm lead}$. For $\gamma_{\rm lead}/t=1$, the distribution is well approximated by Eq.~\eqref{eq.fneq.ballistic.app}, whereas for $\gamma_{\rm lead}/t=0.5$ and $1.5$, strong oscillations appear in the distribution function. This contrasts sharply with the behavior in the diffusive regime shown in Fig.~\ref{fig.fw.lead.x01}(b), where the distribution function remains largely insensitive to $\gamma_{\rm lead}$.

In our model, the parameter $\gamma_{\rm lead}$ in Eq.~\eqref{eq.gam.lead} controls the potential barrier between the wire and the leads. Due to the difference in band structures between the wire and the reservoirs, reflection inevitably occurs at the interfaces. This reflection gives rise to Fabry–Perot-like interference, which results in the oscillations in the weight function shown in Fig.~\ref{fig.wj}(a), as well as in the distribution function shown in Fig.~\ref{fig.fw.lead.x01}(a). As discussed in Appendix~\ref{sec.app.gamma.lead}, the reflection at the interface is minimized at $\gamma_{\rm lead}/t=1$, while it becomes more pronounced as $\gamma_{\rm lead}$ deviates from this value. This is why, as seen in Fig.~\ref{fig.fw.lead.x01}(a), stronger oscillations appear for $\gamma_{\rm lead}/t = 0.5$ and $1.5$ compared to $\gamma_{\rm lead}/t=1$. (We note that, as shown in Appendix~\ref{sec.app.gamma.lead}, the oscillation frequency depends on the system size $N$.) In contrast, in the diffusive regime, since impurity scattering inside the wire dominates over interface scattering, the distribution function becomes almost independent of the interface details.

%%%%%%%%%%%%%%%%%%%%%%%%%%%%%%%%
%%%%%%%%%%%%%%%%%%%%%%%%%%%%%%%%

To summarize, in the presence of elastic scattering from impurities, the distribution function $f^{\rm neq}_j(\omega)$ depends on position $x_j$ while maintaining its characteristic two-step structure in the low-temperature regime ($T_{\rm env}\ll \mu_{\rm L}-\mu_{\rm R}$). Thus, when the wire length is shorter than the electron inelastic mean free path and inelastic scattering is negligible, the local equilibrium assumption is no longer valid so that effective temperature $T_{\rm eff}$ and chemical potential $\mu_{\rm eff}$ cannot be defined in the entire ballistic-diffusive crossover regime. Furthermore, we have shown that in the ballistic regime, the distribution function is highly sensitive to the wire-reservoir coupling parameter $\gamma_{\rm lead}$, whereas in the diffusive regime it becomes nearly independent of $\gamma_{\rm imp}$ as impurity scattering within the wire dominates over interface scattering.

%%%%%%%%%%%%%%%%%%%%%%%%%%%%%%%%%%%%%%%%%%%%%%%%%%%%%%%%%%%%%%%%%%%%%%%%%%%%%%%%%%%%%%%%%%%%%%%%%%%%%%%%%%%%%%%%
\subsection{Crossover from the Non-Equilibrium to the Local-Equilibrium Regime}

We next discuss how inelastic scattering from phonons affects the form of the distribution function. As shown in Figs.~\ref{fig.fneq}(a), (d), and (g), the distribution function depends on position $x_j$ due to inelastic scattering from phonons. Moreover, Fig.~\ref{fig.phi}(b) shows that the profile of the electrostatic potentials $\varphi_j$ changes from the flat profile to the linear profile connecting the electrochemical potentials $\mu_{\alpha}$ in the reservoirs, with increasing the electron-phonon coupling strength $\gamma_{\rm ph}$. These behaviors are similar to those observed when increasing the impurity scattering strength $\gamma_{\rm imp}$, as discussed in Sec.~\ref{sec.result.imp}.

\begin{figure}[t]
\centering
\includegraphics[width=8.2cm]{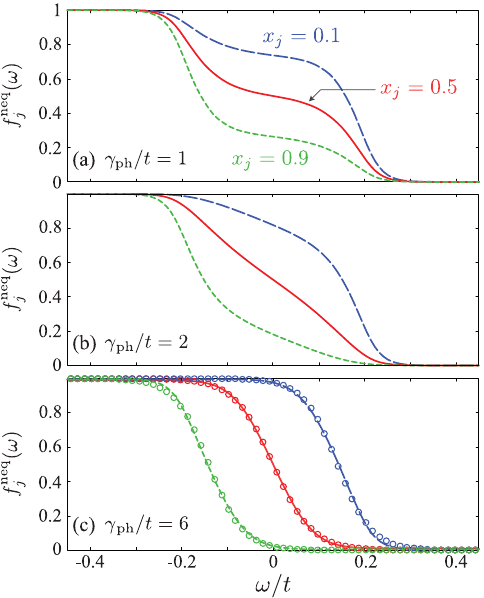}
\caption{Calculated distribution function $f_j^{\rm neq}(\omega)$ for (a) $\gamma_{\rm ph}/t=1$, (b) $\gamma_{\rm ph}/t=2$, and (c) $\gamma_{\rm ph}/t=6$. We set $\gamma_{\rm imp}=0$ for all panels. In panel (c), the calculated distribution function can be well fitted by the Fermi-Dirac distribution in Eq.~\eqref{eq.fneq.hote}, and the fitting results are shown by circles.}
\label{fig.ph}
\end{figure}

In addition to this effect, inelastic scattering from phonons smears out the characteristic two-step structure in the nonequilibrium distribution function. Figure~\ref{fig.ph} shows the distribution function at $x_j=0.1$, $0.5$, and $0.9$ for different electron-phonon coupling strengths $\gamma_{\rm ph}$. As the electron-phonon coupling increases, the nonequilibrium distribution function having the two-step structure gradually evolves into a Fermi-Dirac-like distribution function. We see from Fig.~\ref{fig.ph}(c) that in the presence of strong inelastic scattering from phonons, the distribution function $f_j^{\rm neq}(\omega)$ can be well approximated by the Fermi-Dirac distribution function
\begin{equation}
f_j^{\rm neq}(\omega) \simeq \frac{1}{e^{(\omega -\mu^{\rm eff}_j)/T_j^{\rm eff}}+1},
\label{eq.fneq.hote}
\end{equation}
even under a non-zero bias voltage. In this regime, the electrons reach local equilibrium through inelastic scattering from phonons, which redistributes the electron population distorted by the applied bias~\cite{GueronThesis, heikkilaBook}. Since the distribution function can be well fitted by the Fermi-Dirac distribution function, the effective temperature $T_j^{\rm eff}$ and electrochemical potential $\mu^{\rm eff}_j$ are meaningful physical quantities in this regime.

In Fig.~\ref{fig.ph}(c),  the calculated distribution function $f_j^{\rm neq}(\omega)$ is fitted to Eq.~\eqref{eq.fneq.hote}, yielding fitting parameters $T^{\rm eff}_j/t\simeq 0.043$ and $\mu^{\rm eff}_j/t\simeq 0$ for $x_j=0.5$, and $T^{\rm eff}_j/t\simeq 0.037$ and $\mu^{\rm eff}_j/t \simeq \pm 0.141$ for $x_j=0.1$ $(0.9)$. Although the effective temperature $T^{\rm eff}_j$ deviates from the environment temperature $T_{\rm env}/t=0.02$, the fitting errors remain below 1\% in all cases, indicating that the electronic system has reached a locally thermalized state. If $\gamma_{\rm ph}$ is increased further, the electronic system fully thermalizes with the phonon bath, and the local distribution $f^{\rm neq}_j(\omega)$ is well approximated by
\begin{equation}
f_j^{\rm neq}(\omega) \simeq
\frac{1}{e^{(\omega-e\varphi_j)/T_{\rm env}}+1},
\label{eq.fneq.strong.limit}
\end{equation}
where the effective electronic temperature $T^{\rm eff}_j$ and electrochemical potential $\mu^{\rm eff}_j$ coincide with the phonon bath temperature $T_{\rm env}$ and the local electrostatic potential $e\varphi_j$, respectively~\cite{GueronThesis, heikkilaBook}. This fully thermalized behavior in the strong electron-phonon coupling limit is demonstrated in Appendix~\ref{sec.app.strong.e.ph}.

%%%%%%%%%%%%%%%%%%%%%%%%%%%%%%%%
%%%%%%%%%%%%%%%%%%%%%%%%%%%%%%%%

We note that while electron-phonon interactions are the dominant source of inelastic scattering typically above 1K, electron-electron interactions become the leading inelastic process at lower temperatures~\cite{Altshuler1982, Pierre2003, EfrosBook, heikkilaBook}. Electron-electron scattering smears out the two-step structure in the distribution function induced by the bias voltage, as does electron-phonon scattering. In particular, when the wire length is sufficiently long compared to the electron-electron mean free path, electrons reach local equilibrium through electron-electron scattering~\cite{HuardThesis, GueronThesis, PierreThesis, AnneThesis, heikkilaBook}. In this state, commonly referred to as the ``hot-electron state", the distribution function $f_j^{\rm neq}(\omega)$ can be well approximated by the Fermi-Dirac distribution function in Eq.~\eqref{eq.fneq.hote}, which is characterized by position-dependent temperature $T^{\rm eff}_j$ and electrochemical potential $\mu^{\rm eff}_j$~\cite{HuardThesis, GueronThesis, PierreThesis, AnneThesis, heikkilaBook}. As mentioned in Sec.~\ref{sec.model}, addressing the effects of electron-electron interactions on the distribution function lies beyond the scope of this work. Incorporating these correlation effects into our scheme and investigating the distribution function in the hot-electron regime remains an important challenge.

%%%%%%%%%%%%%%%%%%%%%%%%%%%%%%%%%%%%%%%%%%%%%%%
\section{Summary}

In summary, we have developed a theoretical framework to describe the nonequilibrium distribution function in a nanowire connected between two electrodes with different electrochemical potentials. The voltage-biased wire was modeled as a one-dimensional tight-binding model connected to equilibrium reservoirs with different electrochemical potentials at both ends. We calculated the nonequilibrium distribution function in the wire using the nonequilibrium Green's function technique. For electron scattering processes in the wire, we considered both elastic scattering from impurities and inelastic scattering from phonons within the self-consistent Born approximation.

We have demonstrated that the nonequilibrium distribution functions in various regimes are well described within our framework. In the ballistic regime, where electron scattering in the wire is negligible, the distribution function is spatially uniform and is given by the simple average of the Fermi-Dirac distribution functions in both reservoirs. In the diffusive regime dominated by elastic scattering from impurities, the distribution function linearly interpolates between the Fermi-Dirac distribution functions in the reservoirs at every energy level. Moreover, in the local equilibrium regime with strong electron-phonon scattering, electrons thermalize with phonons, and the distribution function can be well approximated by the Fermi-Dirac distribution function characterized by the phonon temperature and the local electrostatic potential.

We have also calculated the electrostatic potential along the wire. In the ballistic regime, the potential is constant along the wire and the voltage drops only at the contacts between the wire and the reservoirs. On the other hand, when electron-impurity or electron-phonon scattering is present, the voltage drops in the bulk of the wire, resulting in a linear profile connecting the electrochemical potentials in both reservoirs.

We end by listing some future problems. In this work, we focused on a strictly one-dimensional tight-binding model. An important direction for future research is to extend the present framework to multichannel tight-binding models that incorporate transverse degrees of freedom. Such an extension would allow us to analyze transport phenomena that intrinsically rely on multichannel physics, such as the universal $1/3$ Fano factor discussed in diffusive conductors~\cite{Nagaev1992, Beenakker1992, Beenakker1997, Blanter2000}. Although the extension of our formalism to multichannel systems is straightforward, the associated computational cost grows rapidly with the number of channels. Developing more efficient numerical algorithms that can handle multichannel models therefore represents an essential next step.

Another interesting direction is to apply the present scheme to the calculation of the local noise measured in scanning tunneling microscopy setups~\cite{Tikhonov2020}. The local noise directly probes the structure of the nonequilibrium electron distribution at a given position. Extending our scheme to analyze such local noise signals would provide further insight into nonequilibrium electronic states.

Another promising avenue for future research is to apply our scheme to broader classes of nonequilibrium systems. For example, by combining with the Nambu Green's function technique~\cite{Nambu1960}, our scheme can be applied to superconducting heterostructures, such as a voltage-biased normal-metal wire between superconducting electrodes~\cite{Pierre2001} and a voltage-biased superconducting wire between normal-metal electrodes~\cite{Keizer2006, Vercruyssen2012, Seja2021, Hubler2010, Arutyunov2011, Yagi2006, Takane2006, Takane2007, Takane2009}. It can also be applied to periodically driven systems, such as a metal wire under ac voltage~\cite{Shytov2005, Gabelli2013} and electron gases exposed to time-periodic electric field~\cite{Matsyshyn2023, Shi2024}, by combining with the Floquet Green's function technique~\cite{Aoki2014}. In these systems, nonequilibrium distribution functions having characteristic structures can give rise to a variety of exotic quantum many-body phenomena that have not been observed in systems in (local) equilibrium. Exploring such nonequilibrium phenomena is currently one of the most exciting challenges in condensed matter physics, and our scheme would contribute to the further development of this research field.

%%%%%%%%%%%%%%%%%%%%%%%%%%%%%%%
\begin{acknowledgments}
We gratefully thank H. Pothier for useful comments and for drawing our attention to relevant references. We also thank K. Yoshimi, S. Sumita, and Y. Ohashi for stimulating discussions. T.K. was supported by MEXT and JSPS KAKENHI Grant-in-Aid for JSPS fellows Grant No. JP24KJ0055 and No. 25K23363. Y.K. was supported by JSPS KAKENHI No. JP21H01032. This research was also supported by Joint Research by the Institute for Molecular Science (IMS program No. 23IMS1101) (Y.K.) Some of the computations in this work were done using the facilities of the Supercomputer Center, the Institute for Solid State Physics, the University of Tokyo.
\end{acknowledgments}

\appendix
\section{Numerical Implementation}

\subsection{Workflow of the Numerical Calculation \label{sec.app.workflow}}

%%%%%%%%%%%%%%%%%%%%%%%%%%%%%%%%
\begin{figure}[t]
\centering
\includegraphics[width=8.2cm]{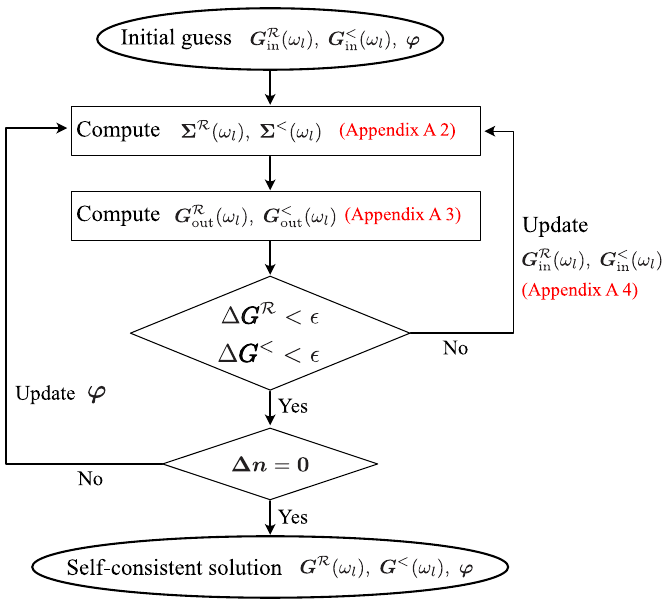}
\caption{Workflow of the numerical calculation used to obtain the self-consistent distribution function $f_j^{\rm neq}(\omega)$, as well as the electrostatic potential $\bm{\varphi}=(\varphi_1,\cdots,\varphi_N)$. Starting from an initial guess for the Green's functions $\bm{G}^{\m{R},<}_{\rm in}$ and the electrostatic potential $\bm{\varphi}$, the impurity and phonon self-energies are computed, followed by the update of the Green's functions. The process is iterated until convergence in the Green’s functions is achieved. Once converged, the local charge densities are computed and used to update $\bm{\varphi}$, ensuring charge neutrality in Eq.~\eqref{eq.null.charge}. The details of each step are explained in Appendix~\ref{sec.app.KK}-\ref{sec.app.Pulay}.}
\label{fig.workflow}
\end{figure}
%%%%%%%%%%%%%%%%%%%%%%%%%%%%%%%%

Figure~\ref{fig.workflow} illustrates the workflow of the numerical calculation for the distribution function $f_j^{\rm neq}(\omega)$. The frequency axis is discretized into a finite set $\{\omega_l\}$ over a certain energy window, and the electrostatic potential is denoted by $\bm{\varphi}=\big(\varphi_1,\cdots,\varphi_N\big)$.

Since the impurity and phonon self-energies, $\bm{\Sigma}^{{\rm X}=\m{R},<}_{\rm imp}$ and $\bm{\Sigma}^{\rm X}_{\rm ph}$, depend on the dressed Green’s functions $\bm{G}^{\rm X}$, we need to determine $\bm{G}^{\rm X}$ self-consistently. The initial guess $\bm{G}^{\rm X}_{\rm in}$ for the Green's functions can be taken either from the ballistic limit ($\gamma_{\rm imp}=\gamma_{\rm ph}=0$) result or the converged result at nearby parameter values.

In evaluating the self-energy corrections from $\bm{G}^{{\rm X}}_{\rm in}$, the most computationally demanding part is the calculation of the retarded phonon self-energy $\bm{\Sigma}^{\m{R}}_{\rm ph}$ in Eq.~\eqref{eq.self.ph.r}. An efficient method for this calculation is presented in Appendix~\ref{sec.app.KK}. Once the all self-energies are evaluated, we compute the updated Green’s functions $\bm{G}^{\rm X}_{\rm out}$. This step corresponds to the inversion of a tridiagonal matrix, as described in Eq.~\eqref{eq.Gr.full}. The efficient procedure for performing this matrix inversion is explained in Appendix~\ref{sec.app.Inv.Tri}.

The iterative process is considered converged when the update error
\begin{equation}
\Delta G^{{\rm X}=\m{R},<}
\equiv
\sum_l \sum_{j,k=1}^{N}
\left|
G^{\rm X}_{{\rm out},jk}(\omega_l)-
G^{\rm X}_{{\rm in},jk}(\omega_l)
\right|
\end{equation}
falls below a specified tolerance $\epsilon$. In this study, we use $\epsilon/t = 10^{-12}$. The convergence rate of this iterative solver directly affects the overall computational cost. To accelerate convergence, we employ the restarted Pulay mixing algorithm, which is detailed in Appendix~\ref{sec.app.Pulay}.

Once a self-consistent solution for the lesser Green's function $\bm{G}^<$ is obtained for a given scalar potential $\bm{\varphi}$, $\bm{\Delta n}=(\Delta n_1, \cdots, \Delta n_N)$ in Eq.~\eqref{eq.null.charge} can be computed.  The entire process is then iterated by updating $\bm{\varphi}$ until the charge neutrality condition $\bm{\Delta n}(\bm{\varphi}) = \bm{0}$ is satisfied. Since this condition can be regarded as a set of nonlinear equations with respect to $\bm{\varphi}$, we employ the Broyden method~\cite{Broyden1965, Broyden1967} to efficiently update $\bm{\varphi}$ during the iteration.

%%%%%%%%%%%%%%%%%%%%%%%%%%%%%%%%%%%%%%%%%%%%%%%%%%%%%%%%%%%%%%%%%%
\subsection{Numerical Hilbert Transformation \label{sec.app.KK}}

To efficiently evaluate $\bm{\Sigma}^{\m{R}}_{\rm ph}(\omega)$ in Eq.~\eqref{eq.self.ph.r}, we take advantage of the fact that the imaginary part ${\rm Im}\bm{\Sigma}^{\m{R}}_{\rm ph}(\omega)$ of the self-energy is obtained from the lesser and the greater components $\bm{\Sigma}^\lessgtr_{\rm ph}(\omega)$ in Eq.~\eqref{eq.self.ph.<.2} as~\cite{RammerBook}
\begin{equation}
{\rm Im}\bm{\Sigma}^{\m{R}}_{\rm ph}(\omega) = \frac{1}{2i}\big[\bm{\Sigma}^>_{\rm ph}(\omega) -\bm{\Sigma}^<_{\rm ph}(\omega)\big].
\end{equation}
The real part is then evaluated from the Kramers-Kronig relation, given by
\begin{equation}
{\rm Re}\bm{\Sigma}^{\m{R}}_{\rm ph}(\omega) = \frac{1}{\pi} \m{P}\int_{-\infty}^\infty d\omega' \frac{{\rm Im}\bm{\Sigma}^{\m{R}}_{\rm ph}(\omega')}{\omega' -\omega}.
\label{eq.app.KK}
\end{equation}
Here, $\m{P}$  denotes the Cauchy principal value integral.

The direct computation of the Hilbert transformation in Eq.~\eqref{eq.app.KK} is typically a numerically demanding task. To circumvent this difficulty, we employ the interpolation technique~\cite{Frederiksen2007, Vaitkus2022}: We approximate the function ${\rm Im}\bm{\Sigma}^{\m{R}}_{\rm ph}(\omega)$ by a linear interpolation to the values ${\rm Im}\bm{\Sigma}^{\m{R}}_{\rm ph}(\omega_l)$ known at discrete grid points $\{\omega_l\}$, expressed as
\begin{equation}
{\rm Im}\bm{\Sigma}^{\m{R}}_{\rm ph}(\omega) \simeq \sum_l {\rm Im}\bm{\Sigma}^{\m{R}}_{\rm ph}(\omega_l) \Phi_l(\omega).
\end{equation}
Here, $\Phi_j(\omega)$ is the kernel function associated with the linear interpolation, given by
\begin{align}
\Phi_l(\omega)
&=
\dfrac{\omega -\omega_{l-1}}{\omega_l -\omega_{l-1}}
\big[\Theta(\omega_l -\omega) -\Theta(\omega_{l-1} -\omega) \big]
\notag\\
&\hspace{0.3cm}
+
\dfrac{\omega_{l+1}-\omega}{\omega_{l+1} -\omega_l}
\big[\Theta(\omega_{l+1} -\omega) -\Theta(\omega_l -\omega) \big].
\end{align}
Substituting the approximated ${\rm Im}\bm{\Sigma}^{\m{R}}_{\rm ph}(\omega)$ into Eq.~\eqref{eq.app.KK}, we have
\begin{equation}
{\rm Re}\bm{\Sigma}^{\m{R}}_{\rm ph}(\omega) =
\sum_l {\rm Im}\bm{\Sigma}^{\m{R}}_{\rm ph}(\omega_l) \phi_l(\omega).
\label{eq.app.HT.simple}
\end{equation}
Here,
\begin{align}
\phi_l(\omega)
&=
\frac{1}{\pi}\m{P}\int_{-\infty}^\infty d\omega'\frac{\Phi_l(\omega')}{\omega' -\omega}	
\notag\\
&=
\frac{1}{\pi}
\Bigg[
\frac{\omega -\omega_{l-1}}{\omega_j -\omega_{l-1}}
\log\left|\frac{\omega_{l-1} -\omega}{\omega_l -\omega}\right|
\notag\\
&\hspace{1.8cm}
+
\frac{\omega -\omega_{l+1}}{\omega_j -\omega_{l+1}}
\log\left|\frac{\omega -\omega_l}{\omega -\omega_{l+1}}\right|
\Bigg]
\end{align}
is the transformation kernel~\cite{Frederiksen2007, Vaitkus2022}. With this kernel, the Hilbert transformation in Eq.~\eqref{eq.app.KK} can be performed by the simple summation in Eq.~\eqref{eq.app.HT.simple}.

%%%%%%%%%%%%%%%%%%%%%%%%%%%%%%%%%%%%%%%%%%%%%%%%%%%%%%%%%%%%%%%%%%%%%%%%%
\subsection{Inverse of a Tridiagonal Matrix \label{sec.app.Inv.Tri}}
The inverse of a non-singular tridiagonal matrix
\begin{equation}
\bm{T}=
\begin{pmatrix}
a_1 & b_1 &  \\
c_1 & a_2 & b_2 \\
& \ddots & \ddots & \ddots  \\
&& c_{n-2} & a_{n-1} & b_{n-1} \\
&&& c_{n-1} & a_n
\end{pmatrix}
\end{equation}
is given by~\cite{Usmani1994}
\begin{equation}
\Big(T^{-1}\Big)_{ij}=
\begin{cases}
(-1)^{i+j}b_i \cdots b_{j-1} \theta_{i-1}\phi_{j+1}/\theta_n
\hspace{0.38cm}(i<j)\\[4pt]
\theta_{i-1}\phi_{j+1}/\theta_n 
\hspace{3cm}(i=j)\\[4pt]
(-1)^{i+j} c_j \cdots c_{i-1} \theta_{j-1}\phi_{i+1}/\theta_n
\hspace{0.38cm}(i>j)
\end{cases},
\end{equation}
where $\theta_i$ satisfies the recurrence relation
\begin{equation}
\theta_i = a_i \theta_{i-1} -b_{i-1}c_{i-1}\theta_{i-2}
\hspace{0.5cm}
(2\leq i \leq n)
\end{equation}
with initial conditions $\theta_0 =1$, $\theta_1 =a_1$. For $\phi_j$, we have
\begin{equation}
\phi_i = a_i \phi_{i+1} -b_i c_i \phi_{i+2}
\hspace{0.5cm}
(n-1\geq i \geq 1)
\end{equation}
with initial conditions $\phi_{n+1}=1$ and $\phi_n=a_n$.

%%%%%%%%%%%%%%%%%%%%%%%%%%%%%%%%
\begin{figure}[t]
\centering
\includegraphics[width=8.2cm]{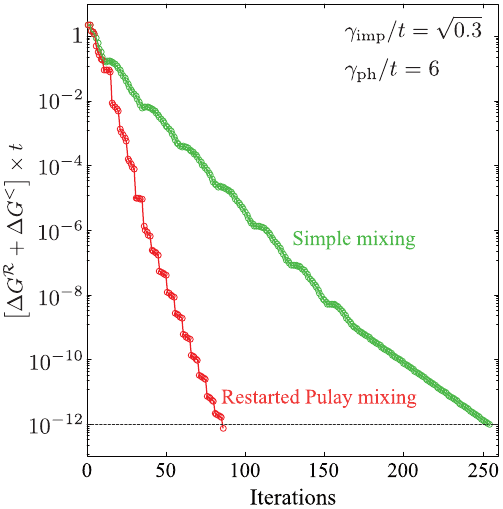}
\caption{Convergence behavior of the self-consistent calculation using the simple mixing in Eq.~\eqref{eq.app.simple.mixing} and the restarted Pulay mixing in Eq.~\eqref{eq.app.Pulay.mixing}. The vertical axis shows the residual norm as a function of iteration number. The restarted Pulay mixing scheme (red) accelerates convergence compared to simple mixing (green). We set $\gamma_{\rm imp}/t=\sqrt{0.3}$, $\gamma_{\rm ph}/t=6$, $\alpha=0.5$, and $m=5$.}
\label{fig.mixing}
\end{figure}
%%%%%%%%%%%%%%%%%%%%%%%%%%%%%%%%

%%%%%%%%%%%%%%%%%%%%%%%%%%%%%%%%
\subsection{Convergence Acceleration \label{sec.app.Pulay}}

A common approach to solving the self-consistency loop is the simple mixing scheme, where the input Green's function $\bm{G}^{\rm X}_{{\rm in},n+1}$ for the next iteration is constructed as a linear combination of the current input and output:
\begin{equation}
\bm{G}^{\rm X}_{{\rm in},n+1}(\omega_l) 
=
\bm{G}^{\rm X}_{{\rm in},n}(\omega_l)
+
\alpha \big[
\bm{G}^{\rm X}_{{\rm out},n}(\omega_l) -
\bm{G}^{\rm X}_{{\rm in},n}(\omega_l) 
\big],
\label{eq.app.simple.mixing}
\end{equation}
where $\alpha$ is a mixing parameter between 0 and 1. While straightforward to implement, this method often suffers from slow convergence, especially when $\gamma_{\rm imp}$ and $\gamma_{\rm ph}$ are large.

To accelerate convergence, we adopt the restarted Pulay mixing scheme~\cite{Pulay1980, Phanisri2015, Amartya2016}, a modification of the direct inversion in the iterative subspace method. In this approach, the next input is constructed as
\begin{equation}
\bm{G}^{\rm X}_{{\rm in},n+1}(\omega_l) 
=
\bar{\bm{G}}^{\rm X}_{{\rm in},n}(\omega_l)
+
\alpha \bar{\bm{r}}^{\rm X}_n(\omega_l),
\label{eq.app.Pulay.mixing}
\end{equation}
where $\bar{\bm{G}}^{\rm X}_{{\rm in},n}$ and $\bar{\bm{r}}^{\rm X}_n$ are optimized averages over the previous $(m+1)$ iterates and residuals:
\begin{subequations} \label{all}
\begin{align}
&
\bar{\bm{G}}^{\rm X}_{{\rm in},n}(\omega_l)=
\bm{G}^{\rm X}_{{\rm in},n}(\omega_l) 
-
\sum_{j=1}^m c_j \Delta \bm{G}^{\rm X}_{{\rm in},n-m+j}(\omega_l), 
\\
&
\bar{\bm{r}}^{\rm X}_n(\omega_l)=
\bm{r}^{\rm X}_n(\omega_l)
-
\sum_{j=1}^m c_j \Delta \bm{r}^{\rm X}_{n-m+j}(\omega_l).
\end{align}
\end{subequations}
Here, the differences are defined as
\begin{subequations} \label{all}
\begin{align}
&
\Delta \bm{G}^{\rm X}_{{\rm in},n}(\omega_l) = \bm{G}^{\rm X}_{{\rm in},n}(\omega_l) - \bm{G}^{\rm X}_{{\rm in},n-1}(\omega_l)
,\\
&
\Delta \bm{r}^{\rm X}_n(\omega_l) = \bm{r}^{\rm X}_n(\omega_l) - \bm{r}^{\rm X}_{n-1}(\omega_l)
,\\
&
\bm{r}^{\rm X}_n(\omega_l) = \bm{G}^{\rm X}_{{\rm out},n}(\omega_l) -\bm{G}^{\rm X}_{{\rm in},n}(\omega_l).
\end{align}
\end{subequations}
The scalar coefficients \(\bm{c} = (c_1, \cdots, c_m)^{\rm T}\) are determined by minimizing the Frobenius norm of the averaged residual:
\begin{equation}
\bm{c} = \operatorname*{arg\,min}_{\bm{c}} \|\bar{\bm{r}}^{\rm X}_n(\omega_l)\|_{\rm F}.
\label{eq.def.c}
\end{equation}
It is shown that this least-squares problem leads to the linear equation~\cite{Fang2009}
\begin{equation}
\left[
\left(\Delta \bm{R}^{\rm X}_n\right)^{\rm T}
\Delta \bm{R}^{\rm X}_n
\right]
\bm{c}
=
\left(\Delta \bm{R}^{\rm X}_n\right)^{\rm T}
\bm{r}^{\rm X}_n,
\label{eq.linear.for.c}
\end{equation}
where
\begin{equation}
\Delta \bm{R}^{\rm X}_n = \left( \Delta \bm{r}^{\rm X}_{n-m+1}, \Delta \bm{r}^{\rm X}_{n-m+2}, \cdots, \Delta \bm{r}^{\rm X}_{n} \right).
\end{equation}

To enhance numerical stability and avoid the accumulation of nearly linearly dependent vectors, the restarted Pulay scheme resets the history after every $(m+1)$ iterations. In our implementation, we typically use a history length of $m=5$. This strategy significantly reduces the number of iterations required for convergence, especially under strong elastic and inelastic scattering conditions, as demonstrated in Fig.~\ref{fig.mixing}. Since the additional computational cost per iteration is negligible, as it only involves solving the small linear system in Eq.~\eqref{eq.linear.for.c}, the reduction in the number of iterations directly decreases in the total computational cost. Moreover, as illustrated in Fig.~\ref{fig.workflow}, because the self-energy iteration is repeated multiple times to determine the electrostatic potential, the accelerated convergence further results in a substantial overall reduction in runtime.

%%%%%%%%%%%%%%%%%%%%%%%%%%%%%%%%%%%%%%%%%%%%%%%
\section{Electron Transport Properties in the Ballistic-Diffusive Crossover Regime \label{sec.app.transport}}

The charge current $I_{\alpha={\rm L}, {\rm R}}(t)$ from the $\alpha$ reservoir to the wire is determined from the rate of change in the number of electrons in the $\alpha$ reservoirs~\cite{Meir1992, Jauho1994, HaugBook}:
\begin{align}
I_\alpha(t) 
&= -e \frac{d}{dt} \sum_{\bm{k}} \braket{a^\dagger_{\alpha,\bm{k}}(t) a_{\alpha, \bm{k}}(t)}
\notag\\
&=
-2e {\rm Re} \sum_{\bm{k}} \Big[t_{\rm lead} G^<_{{\rm mix},\alpha,\bm{k}}(t,t)\Big].
\label{eq.current.def}
\end{align}
Here, we have introduced the mixed lesser function $G^<_{{\rm mix},\alpha,\bm{k}}(t,t')$, defined by
\begin{equation}
G^<_{{\rm mix},\alpha,\bm{k}}(t,t')= i \braket{a^\dagger_{\bm{k}}(t') c_{i_\alpha}(t)},
\end{equation}
where $i_\alpha=1$ if $\alpha={\rm L}$ and $i_\alpha=N$ if $\alpha={\rm R}$. When the system is in a NESS, this function is evaluated as~\cite{Meir1992, Jauho1994, HaugBook}
\begin{align}
G^<_{{\rm mix},\alpha,\bm{k}}(t,t)&=
-t_{\rm lead}^* \int_{-\infty}^\infty d\omega \hspace{0.1cm}
\Big[G^{\m{R}}_{i_\alpha i_\alpha}(\omega) \mathscr{G}^<_{\alpha, \bm{k}}(\omega)
\notag\\
&\hspace{0.5cm}
+
G^<_{i_\alpha i_\alpha}(\omega) \mathscr{G}^{\m{A}}_{\alpha, \bm{k}}(\omega)
\Big].	
\label{eq.G<.mix}
\end{align}
Here, $G^{\m{R}, <}_{i_\alpha i_\alpha}(\omega)$ is the dressed Green's function, obtained from the Dyson equations~\eqref{eq.Dyson.r} and \eqref{eq.Dyson.<}, while $\mathscr{G}^{\m{A}, <}_{\alpha, \bm{k}}(\omega)$ represents the noninteracting Green's function in the $\alpha$ reservoir, given in Eqs.~\eqref{eq.nonint.G.RA} and \eqref{eq.nonint.G.<}. Substituting Eq.~\eqref{eq.G<.mix} into Eq.~\eqref{eq.current.def} and performing the $\bm{k}$ summation, we obtain the charge current as
\begin{subequations} \label{all}
\begin{align}
I_\alpha &= 
2e{\rm Re}
\int_{-\infty}^\infty \frac{d\omega}{2\pi} 
\Big[
G^{\m{R}}_{i_\alpha i_\alpha}(\omega) \Sigma^<_{{\rm lead}, \alpha}(\omega)
\notag\\
&\hspace{3.5cm}+
G^<_{i_\alpha i_\alpha}(\omega) \Sigma^{\m{A}}_{{\rm lead}, \alpha}(\omega)
\Big]
\\
&= 
2e {\rm Re} \int_{-\infty}^\infty \frac{d\omega}{2\pi} {\rm Tr}
\Big[
\bm{G}^{\m{R}}(\omega) \bm{\Sigma}^<_{{\rm lead}, \alpha}(\omega)
\notag\\
&\hspace{3.5cm}+
\bm{G}^<(\omega) \bm{\Sigma}^{\m{A}}_{{\rm lead}, \alpha}(\omega)
\Big]
\label{eq.current.2.sub}
\\
&=
2i e
\int_{-\infty}^\infty \frac{d\omega}{2\pi}\hspace{0.1cm}
{\rm Tr}
\bigg[\bm{\Gamma}_\alpha
\Big[
f(\omega-\mu_\alpha) 
\big[\bm{G}^{\m{R}}(\omega)-\bm{G}^{\m{A}}(\omega)\big]
\notag\\
&\hspace{3.5cm}
+
\bm{G}^<(\omega)
\Big]
\bigg].
\label{eq.current.2}	
\end{align}
\end{subequations}
In deriving Eq.~\eqref{eq.current.2} form Eq.~\eqref{eq.current.2.sub}, we have used Eqs.~\eqref{eq.self.lead.RA} and \eqref{eq.self.lead.<}. Noting that $I = I_{\rm L}= -I_{\rm R}$ in the NESS, we obtain a symmetric expression for the current, known as the Meir-Wingreen formula~\cite{Meir1992}, as
\begin{align}
I
&=
ie 
\int_{-\infty}^\infty \frac{d\omega}{2\pi}\hspace{0.1cm}
{\rm Tr}
\bigg[
\big[
\bm{\Gamma}_{\rm L}f(\omega-\mu_{\rm L}) -
\bm{\Gamma}_{\rm R}f(\omega-\mu_{\rm R})
\big] 
\notag\\
&\hspace{0.5cm}\times
\big[\bm{G}^{\m{R}}(\omega)-\bm{G}^{\m{A}}(\omega)\big]
+
\big[\bm{\Gamma}_{\rm L}-\bm{\Gamma}_{\rm R}\big]
\bm{G}^<(\omega)
\bigg].
\label{eq.sym.current}
\end{align}

%%%%%%%%%%%%%%%%%%%%%%%%%%%%%%%%
\begin{figure}[t]
\centering
\includegraphics[width=8.2cm]{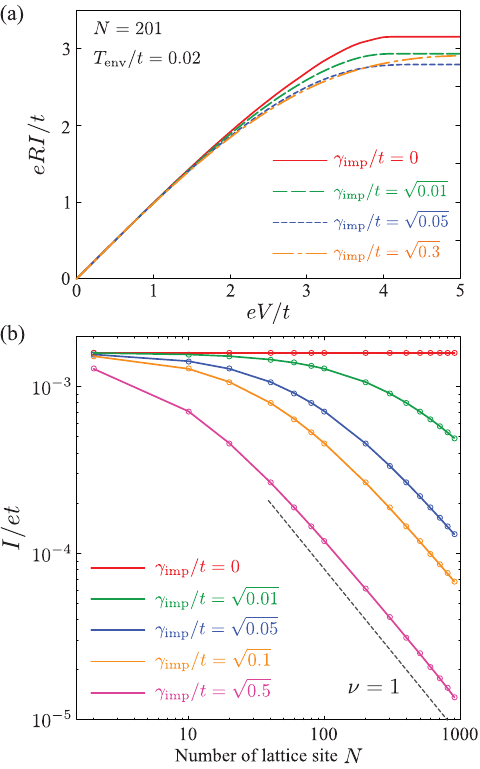}
\caption{(a) Voltage-current characteristics of the wire and (b) system size (number of lattice site $N$) dependence of the current through the wire. We show the results for different values of impurity scattering strength $\gamma_{\rm imp}$. In panel (a), the current is normalized with the linear resistance $R$. In panel (b), the bias voltage is set to $eV/t=0.01$.}
\label{fig.current}
\end{figure}
%%%%%%%%%%%%%%%%%%%%%%%%%%%%%%%%

We note that in the ballistic limit ($\gamma_{\rm imp}=\gamma_{\rm ph}=0$), Eq.~\eqref{eq.sym.current} can be simplified by using Eq.~\eqref{eq.formula.GR.GA}, leading to
\begin{equation}
I = e\int_{-\infty}^\infty \frac{d\omega}{2\pi}\hspace{0.1cm}
\big[f(\omega-\mu_{\rm L}) -f(\omega-\mu_{\rm R}) \big]
T(\omega).
\label{eq.Landauer}
\end{equation}
Here,
\begin{equation}
T(\omega)=4\gamma_{\rm lead}^2 |G^{\m{R}}_{1N}(\omega)|^2
\label{eq.transmission.Tw}
\end{equation}
represents the transmission probability of the ballistic wire. Equation~\eqref{eq.Landauer} is known as the Landauer formula~\cite{Landauer1957, Landauer1970}. We emphasize that  Eq.~\eqref{eq.Landauer} is valid only in the ballistic limit.

Figure~\ref{fig.current}(a) presents the voltage-current characteristics of the wire, computed from Eq.~\eqref{eq.sym.current}. In this figure, the current $I$ is normalized by $R$, which is defined as the resistance in the linear response regime, $R\equiv \left.V/I\right|_{V\to 0}$. This result indicates that the bias voltage $eV/t = 0.4$, used in Fig.~\ref{fig.fneq}, lies in the linear transport regime ($I\propto V$) in the entire ballistic-diffusive crossover regime. We note that the current $I$ saturates when the applied bias voltage $V$ exceeds the bandwidth $W=4t$ of the wire.

In the linear transport regime, the current $I$ follows a power-law scaling with the system size (number of lattice sites $N$)~\cite{StegmannThesis, Jin2022}:
\begin{equation}
I \sim \frac{1}{N^\nu}.
\end{equation}
As shown in Fig.~\ref{fig.current}(b), the current remains independent of the system size ($\nu=0$) in the ballistic limit ($\gamma_{\rm imp}=0$). In contrast, in the presence of impurity scattering ($\gamma_{\rm imp}\neq 0$), the current depends on the system size $N$. In particular, for $\gamma_{\rm imp}/t = \sqrt{0.5}$, the current is inversely proportional to the system size ($\nu=1$), which is a characteristic feature of an Ohmic conductor. These changes in the ballistic-diffusive crossover regime are consistent with results from the dephasing model~\cite{Datta1989, McLennan1991, Golizadeh2007, Jin2022} and self-consistent reservoir model~\cite{Amato1990, Roy2007}, both of which are widely used to study electron transport in the ballistic-diffusive crossover regime.

%%%%%%%%%%%%%%%%%%%%%%%%%%%%%%%%%%%%%%%%%%%%%%%
\section{$\gamma_{\rm lead}$ Dependence of the Interface Potential Barrier \label{sec.app.gamma.lead}}

%%%%%%%%%%%%%%%%%%%%%%%%%%%%%%%%
\begin{figure}[t]
\centering
\includegraphics[width=8.2cm]{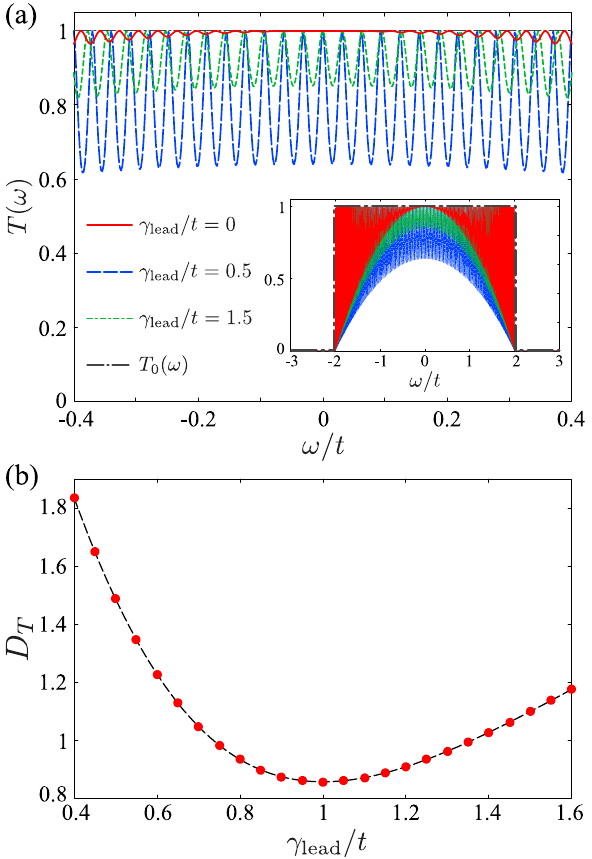}
\caption{(a) Energy dependence of the transmission probability $T(\omega)$ in the ballistic limit for $\gamma_{\rm lead}/t=0.5$, $1$, and $1.5$. The inset shows $T(\omega)$ over a wider energy window, while the main panel focuses on $-0.4<\omega/t<0.4$. The dash-dot line $T_0(\omega)$ represents the ideal case without reflection at the interfaces, where $T_0(\omega)=1$ within the bandwidth and $T_0(\omega)=0$ outside. (b) $\gamma_{\rm lead}$ dependence of the integrated deviation $D_T$ in Eq.~\eqref{eq.app.def.DT}. The minimum of $D_T$ at $\gamma_{\rm lead}/t=1$ indicates that the smoothest interface in this model is realized around this value.}
\label{fig.Tw}
\end{figure}
%%%%%%%%%%%%%%%%%%%%%%%%%%%%%%%%

%%%%%%%%%%%%%%%%%%%%%%%%%%%%%%%%
\begin{figure}[t]
\centering
\includegraphics[width=8.2cm]{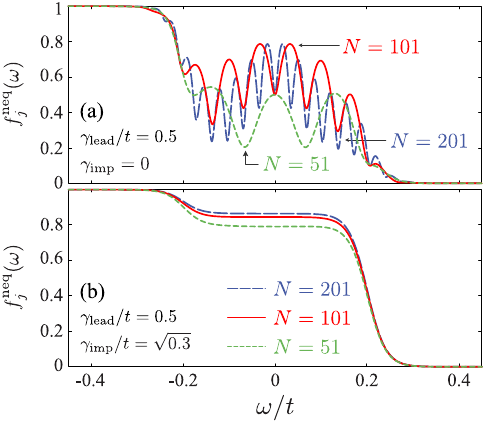}
\caption{Distribution function $f^{\rm neq}_j(\omega)$ at position $x_j=0.1$ for different system sizes $N$ in (a) the ballistic limit ($\gamma_{\rm imp}=0$) and (b) the diffusive regime ($\gamma_{\rm imp}/t=\sqrt{0.3}$). In both panels, we set $\gamma_{\rm lead}/t=0.5$.}
\label{Fig_fw_lead_x01_N}
\end{figure}
%%%%%%%%%%%%%%%%%%%%%%%%%%%%%%%%

In this appendix, we discuss how the potential barrier at the wire-reservoir interface depends on the coupling parameter $\gamma_{\rm lead}$ in Eq.~\eqref{eq.gam.lead}. To this end, we analyze the wire's transmission probability $T(\omega)$ in the ballistic limit, which is given in Eq.~\eqref{eq.transmission.Tw}. In this limit, electron reflection occurs only at the interfaces between the wire and the reservoirs, so that $T(\omega) < 1$ directly reflects the presence of an effective potential barrier at the interfaces.

Figure~\ref{fig.Tw}(a) shows the calculated transmission probability $T(\omega)$ for different values of $\gamma_{\rm lead}$. When no reflection occurs at the interfaces, $T(\omega)=1$ within the wire's bandwidth $(-2<\omega/t<2)$ and $T(\omega)=0$ outside~\cite{ScheerBook, StegmannThesis}, as shown by the dash-dot line $T_0(\omega)$ in Fig.~\ref{fig.Tw}(a). As the potential barrier at the interface increases, the deviation of $T(\omega)$ from $T_0(\omega)$ becomes more pronounced, exhibiting strong oscillatory behavior as a function $\omega$. Figure~\ref{fig.Tw}(b) quantifies this deviation by plotting
\begin{equation}
D_T \equiv \int_{-\infty}^\infty d\omega |T(\omega) -T_0(\omega)|
\label{eq.app.def.DT}
\end{equation}
as a function of $\gamma_{\rm lead}$. The result shows that $D_T$ takes its minimum at $\gamma_{\rm lead}/t=1$, indicating that the smoothest interface in our model is realized around this value. As $\gamma_{\rm lead}/t$ deviates from unity, $D_T$ increases, reflecting the enhanced interface mismatch.

It is important to note that, due to the difference in band structures between the wire and the reservoirs, reflection at the interface inevitably occurs regardless of the value of $\gamma_{\rm lead}$. The backscattering at the interface gives rise to Fabry–Perot-like interference, which manifests as oscillations of $T(\omega)$, as seen in Fig.~\ref{fig.Tw}(a). The oscillation frequency of $T(\omega)$ depends on the system size $N$~\cite{ScheerBook, StegmannThesis}, and this same frequency is observed in the oscillatory structure of the distribution function $f_j^{\rm neq}(\omega)$ shown in Fig.~\ref{fig.fw.lead.x01}(a). Indeed, as shown in Fig.~\ref{Fig_fw_lead_x01_N}(a), the oscillation pattern of the distribution function varies with $N$ in the ballistic limit, reflecting the sensitivity of the coherent interference to the system size $N$. By contrast, in the diffusive regime, the distribution function at $x_j=0.1$ remains essentially unchanged with $N$, as shown in Fig.~\ref{Fig_fw_lead_x01_N}(b). This indicates that in the ballistic regime, the shape of the distribution function is sensitive to both $\gamma_{\rm lead}$ and the system size $N$, whereas in the diffusive regime it becomes largely insensitive to these parameters.

%%%%%%%%%%%%%%%%%%%%%%%%%%%%%%%%%%%%%%%%%%%%%%%
\section{Distribution Function in the Strong Electron-Phonon Coupling Limit \label{sec.app.strong.e.ph}}

%%%%%%%%%%%%%%%%%%%%%%%%%%%%%%%%
\begin{figure}[t]
\centering
\includegraphics[width=8.2cm]{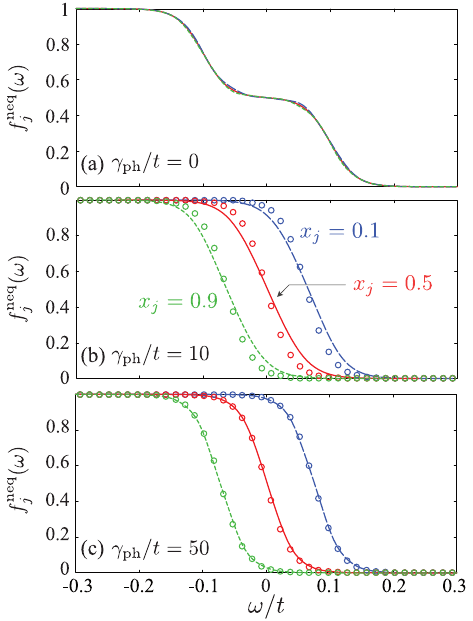}
\caption{Calculated distribution function $f_j^{\rm neq}(\omega)$ for (a) $\gamma_{\rm ph}/t=0$, (b) $\gamma_{\rm ph}/t=10$, and (c) $\gamma_{\rm ph}/t=50$. We set $\gamma_{\rm imp}=0$, $N=101$, $eV/t=0.2$, and $T_{\rm env}/t=0.02$. In panels (b) and (c), circles show the approximated Fermi-Dirac distribution given by Eq.~\eqref{eq.fneq.strong.limit}.}
\label{fig.fneq.strong.limit}
\end{figure}
%%%%%%%%%%%%%%%%%%%%%%%%%%%%%%%%

In the main text, we showed that in the presence of strong inelastic scattering, the local distribution function $f^{\rm neq}_j(\omega)$ can be well approximated by a Fermi-Dirac distribution function with position-dependent effective temperature $T^{\rm eff}_j$ and electrochemical potential $\mu_j^{\rm eff}$, even under a non-zero bias voltage. As the electron-phonon coupling $\gamma_{\rm ph}$ is increased further, the electronic system fully thermalizes with the phonon bath, and the distribution function asymptotically approaches Eq.~\eqref{eq.fneq.strong.limit}. Here, we demonstrate this behavior using a smaller system size $N$ and lower bias voltage $eV$ than those in Fig.~\ref{fig.ph}, which enables us to explore stronger electron–phonon coupling $\gamma_{\rm ph}$ regimes without encountering numerical difficulties.

Figure~\ref{fig.fneq.strong.limit} compares the calculated local distribution function $f^{\rm neq}_j(\omega)$ with the Fermi-Dirac distribution given by Eq.~\eqref{eq.fneq.strong.limit} (circles). Here we set $\gamma_{\rm imp}=0$, $N=101$ and $eV/t=0.2$, while all other parameters are the same as those used in Fig.~\ref{fig.fneq}. As $\gamma_{\rm ph}$ increases, the two-step structure observed at $\gamma_{\rm ph}=0$ disappears, and the local distribution function becomes well approximated by a Fermi-Dirac function, similar to the behavior observed in Fig.~\ref{fig.ph}. As seen in Fig.~\ref{fig.fneq.strong.limit}(b), at $\gamma_{\rm ph}/t=10$, the distribution function already takes the shape of a Fermi–Dirac function; however, a clear deviation from Eq.~\eqref{eq.fneq.strong.limit} is visible, indicating that the effective electronic temperature $T_j^{\rm eff}$ is higher than the phonon bath temperature $T_{\rm env}$. This shows that the electrons are locally thermalized due to inelastic scattering but have not yet fully equilibrated with the phonon bath, which is a state analogous to that shown in Fig.~\ref{fig.ph}(c). On the other hand, at stronger electron-phonon coupling ($\gamma_{\rm ph}/t=50$), the calculated distribution function becomes very well approximated by Eq.~\eqref{eq.fneq.strong.limit}, as shown in Fig.~\ref{fig.fneq.strong.limit}(c). In this strong-coupling regime, the electronic system and phonon bath are fully thermalized, and the effective electronic temperature $T_j^{\rm eff}$ and electrochemical potential $\mu_j^{\rm eff}$ are given by the phonon bath temperature $T_{\rm env}$ and the electrostatic potential $e\varphi_j$, respectively.

These results demonstrate that our model and scheme can consistently capture the entire process in which the nonequilibrium electronic system first attains local thermalization and eventually reaches full equilibrium with the phonon bath, as the electron-phonon coupling $\gamma_{\rm ph}$ increases.

\bibliography{Fneq_ref}
\end{document}